\documentclass[journal]{IEEEtran}
\usepackage[pdftex]{graphicx}
\usepackage[nocompress]{cite}
\usepackage{amsmath}
\usepackage{verbatim}

\begin{document}

\title{Design Framework and Manufacturing of an Active Magnetic Bearing Spindle for Micro-Milling Applications}

\author{Kazi Sher Ahmed
        ~and~Bekir Bediz
\thanks{Kazi Sher Ahmed and Bekir Bediz (corresponding author) are with the Faculty of Engineering and Natural Sciences, Mechatronics Engineering Program, Sabanci University, Istanbul, 34956, Turkiye (email: \{kazisherahmed, bbediz\}@sabanciuniv.edu).}
}


\maketitle

\begin{abstract}
Micro-milling spindles require high rotational speeds where conventional rolling element bearings face limitations such as friction and thermal expansion. Active magnetic bearings (AMBs) address these challenges by providing non-contact and lubrication-free operation at ultra-high speeds with the ability to actively regulate spindle dynamics. The existing literature on AMB spindles has mainly reported specific prototype realizations or control system implementations for specific spindle dynamics. Consequently, design knowledge remains fragmented across isolated successful studies. This paper addresses this gap by presenting a systematic and iterative framework to design and manufacture a micro-milling AMB spindle. The process involves a multidisciplinary design flow with a focus on critical practical aspects of manufacturing. The realized spindle is reported as a case study. 
\end{abstract}


\section{Introduction}\label{sec:intro}

\IEEEPARstart{A}{} primary challenge in mechanical micro-milling is the requirement of relatively high rotational speeds to maintain optimal cutting speed and efficiency with smaller tool diameters when compared with conventional macro-milling~\cite{chaeInvestigationMicrocuttingOperations2006}. Achieving such high rotational speeds poses significant challenges for conventional rolling-element bearings, primarily due to thermal growth, resulting deformation, and increased frictional losses. Non-contact bearings such as air bearings and active magnetic bearings (AMBs) overcome these challenges by enabling non-contact, lubricant-free, high-speed operation with minimal wear. In addition, AMBs provide the advantage to actively regulate the rotor motion, which can be explored to improve process accuracy~\cite{leeAdaptiveControlActive2016} and suppress chatter~\cite{knospeActiveMagneticBearings2007}.

AMB-based machining spindle implementations have been studied in the literature. Kimman \textit{et al.}~\cite{kimmanMiniatureMillingSpindle2010} designed and realized a miniaturized milling spindle with a target speed of 150,000~rpm, while Park \textit{et al.}~\cite{parkMagneticallySuspendedMiniature2012} prototyped a miniature air turbine spindle for tool orbit control. Other implementations focus more intensively on control-oriented challenges; for instance, Knospe~\cite{knospeActiveMagneticBearings2007} demonstrated the active suppression of machining chatter using two AMB test rigs, and Lee and Chen~\cite{leeAdaptiveControlActive2016} presented a novel magnetic actuator and an adaptive control strategy for milling applications. Although these studies demonstrate important capabilities of AMB spindles, they are primarily communicated as application-specific or control-centric implementations rather than a unified design framework encompassing detailed component design and manufacturing. The contribution of this study is the development of a systematic, iterative, and multidisciplinary design framework for micro-milling spindles supported by AMBs with an explicit focus on practical aspects of manufacturing and assembly. The proposed framework enables engineers/designers to navigate the strongly coupled magnetic, mechanical, and thermal challenges of designing an AMB spindle from scratch in a structured and reproducible manner.

This design framework is demonstrated through a detailed case study where a micro-milling AMB spindle is designed and manufactured with a target speed of 110,000~rpm. The case study begins with the identification of application-specific requirements that guide the overall design of the spindle. This is followed by the design, analysis, and realization of the rotational drive, rotor, AMBs, backup bearings, and housings, along with the decisions on instrumentation, control system hardware, and thermal operating limits. Finally, the practical aspects of the spindle assembly are discussed.

\section{Design Framework} \label{sec: SysDesFrame}
\begin{figure}[h!]
  \centering
  \includegraphics[width=8.5cm]{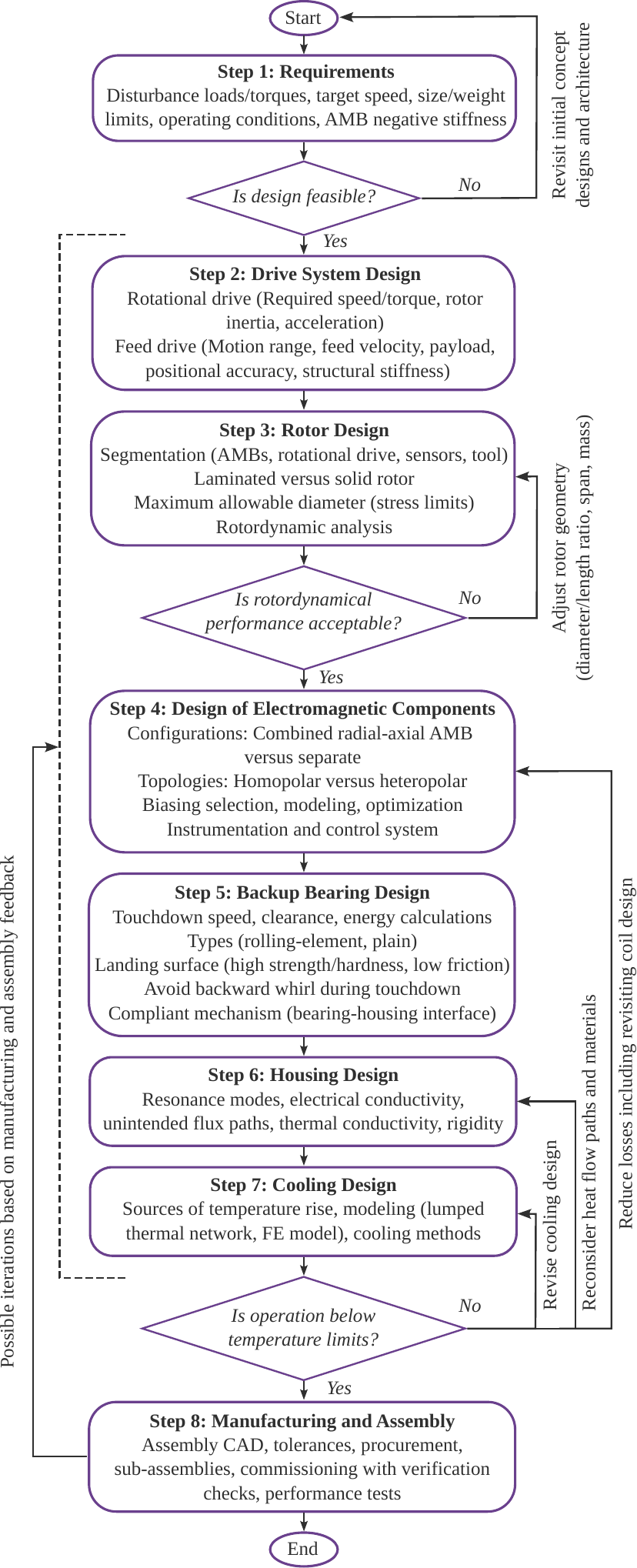}
  \caption{Design flowchart for a micro-milling AMB spindle.}
  \label{fig: flowchart}
\end{figure}

This section presents an eight-step systematic framework to design and manufacture a miniaturized high-speed AMB spindle for micro-milling applications. The multidisciplinary and iterative workflow begins with requirement definition and feasibility assessment and culminates in the manufacturing and assembly of the final system. The proposed framework is illustrated in a flowchart as shown in Figure~\ref{fig: flowchart}.  

\subsection{Step 1: Requirements and Feasibility}  \label{sec: Step1:Req}
The design of a micro-milling AMB spindle starts by defining the application-specific requirements including disturbance loads, (target) rotational speed, dimensional and weight limits, and environmental conditions. A primary task in this step is the evaluation of disturbances (forces and torques) acting on the rotor to size AMB load capacities and spindle rotational drive torque margin. Force disturbances include rotor unbalance and tool cutting forces derived from machining process parameters, while torque loads arise from tangential components of cutting forces and aerodynamic drag on the rotor. A target rotational speed dictated by the required cutting speed and typical micro-milling tool diameters sets the goals for rotordynamical performance and limits the permissible rotor external/outer diameter due to centrifugal stress constraints. In addition, the real-time control system design defines an acceptable range for AMB negative stiffness, which is the open-loop position stiffness induced by bias flux that must be stabilized by the feedback control. 

Based on these requirements, several initial concepts are developed and evaluated through a feasibility study, which acts as a sanity check on the technical and operational viability of the spindle design before proceeding to detailed development~\cite{oleConceptRealityRole2025}. For example, electromagnetic analysis (analytical or FE) can evaluate whether the initial AMB concept can deliver sufficient force against disturbance loads at the target rotational speed while staying within weight and size limits. Initial drive sizing or vendor data can reveal whether the concept rotational drive can deliver an acceptable torque margin. Similarly, a simplified rotordynamic finite-element (FE) model, typically based on beam elements, can produce a Campbell diagram and confirm that flexural resonances lie outside the operating speed range. A preliminary thermal analysis is also performed to guide early decisions regarding coil current density and cooling system.  If the design fails to satisfy feasibility criteria, the initial architecture and key assumptions are revisited before advancing to subsequent steps.

\subsection{Step 2: Drive System Design}    \label{sec: Step2:Drive}
The drive system consists of a spindle rotational drive and a machine feed drive. The selection among potential rotational drives such as electric motors and air (pneumatic) turbines depends on required speed and torque range, rotor inertia, and acceleration requirements. The appropriate feed drives for micro-milling machines include direct-drive electromagnetic stages and indirect rotary-servo stages. The selection should be based on the requirements of the motion range, feed velocity, acceleration capability, payload capacity, positional accuracy, and structural stiffness limits \cite{huoDesignFiveaxisUltraprecision2009}. The selection and design of the rotational drive is tightly coupled to the rotor design, necessitating concurrent consideration of these subsystems.

\subsection{Step 3: Rotor Design}     \label{sec: Step3:Rotor}
The rotor of the spindle is segmented to accommodate the rotational drive and provide segments for radial/axial AMBs, sensors, and tool holder. The rotational speed requirement sets the maximum allowable diameter to keep centrifugal stresses within safe limits. While it is preferred to laminate AMB-facing rotor segments with electrical steel to reduce eddy current losses, the maximum rotational speed determines if laminations can withstand centrifugal stresses through mechanical retention or if a full solid rotor is necessary. Once the preliminary rotor dimensions have been determined, a rotordynamic analysis (via FE method based on Timoshenko beam theory) is conducted to identify critical speeds and verify that flexible resonances are outside of the operating speed range with an adequate separation margin. If rotordynamic requirements are not satisfied, iterative adjustments are made to rotor diameter-to-length ratios, bearing span, and mass distribution until acceptable dynamic performance is achieved.

\subsection{Step 4: Design of Electromagnetic Components} \label{sec: Step4:EM}
Following rotor design, the electromagnetic components of the spindle (radial and axial AMBs) are designed. In cases where both radial and axial rotor motions are controlled by AMBs, two design configurations can be explored: \textbf{(i)} a combined radial-axial AMB, and \textbf{(ii)} separate radial and axial AMBs. A combined AMB requires a relatively shorter rotor but increases the design complexity in terms of electromagnetic cross-coupling \cite{filatovHomopolarPermanentmagnetbiasedActuators2016}. In parallel, AMB topology selection is considered. High-speed radial AMB designs commonly employ homopolar AMB topologies due to their reduced hysteresis and eddy current losses in rotor compared to heteropolar topologies. Bias flux generation is another key design choice, with bias current and permanent magnet (PM) biasing representing the two primary approaches. Current-biasing allows tunable bias current levels but leads to continuous copper loss. In contrast, PM-biasing reduces power consumption by eliminating DC bias current requirement but increases mechanical and assembly complexity \cite{flemingMagneticBearingsStateArt}. These selections in configuration, topology, and biasing depend on application requirements.

Equivalent magnetic circuit models are well-suited for initial AMB design and are computationally inexpensive for optimization studies~\cite{ahmedIntegratedDesignOptimization2025}. However, an FE model is necessary to determine the correction factors to account for simplifications assumed in the analytical models. The final AMB design must provide sufficient static and dynamic load capacity with appropriate safety margins.

Once the AMB topology is finalized, decisions about instrumentation can be made. Here, the requirements of measurement range, sensor bandwidth, resolution, temperature rating, and magnetic field sensitivity are determined. For example, sensor bandwidth puts an upper limit on the frequency up to which an accurate position measurement can be made, 
as attenuation and phase lag become significant once vibration frequencies approach the frequency response limit. Sensor resolution also puts an upper limit on the final achievable positional accuracy of the spindle~\cite{kimmanDesignMicroMilling2010}. Suitable power amplifiers and real-time hardware are selected as well.

\subsection{Step 5: Backup (Auxiliary) Bearing Design}  \label{sec: Step5:BackupBear}
On loss of levitation control due to component failure or overload, the high-speed rotor can cause catastrophic damage to machine components such as stator poles and sensors. Backup (auxiliary or touchdown) bearings as standard subsystems in AMBs provide a fail-safe approach to catch the rotor before it can strike other components. The design usually starts by defining touchdown conditions such as rotational speed at loss of levitation, touchdown clearance, and energy to be absorbed. Commonly used bearing types include rolling-element and plain bearings. The landing interface must have high strength, high hardness, and low friction. A high-friction contact between rotor and touchdown bearing can develop into a destructive backward whirl and contact forces can become as high as 300 times the rotor weight~\cite{schweitzerSafetyReliabilityAspects2005}.  For rolling-element bearings, coated or ceramic rolling elements have demonstrated favorable performance, while cageless designs are often employed in high-speed applications to reduce inertial effects during touchdown~\cite{schweitzerMagneticBearingsTheory2009}. In plain bearing types, ceramic or hard and low-friction materials have been used.

To mitigate impact forces and suppress undesirable dynamic behavior during touchdown, the interface between the backup bearing and the housing is frequently implemented using a compliant support. Solutions such as elastomer O-rings and ribbon dampers have been shown to provide effective impact attenuation while preserving alignment~\cite{jarrouxInvestigationsDynamicBehaviour2021}.

\subsection{Step 6: Housing Design}   \label{sec: Step6:Housing}
Housings are support structures to ensure the proper arrangement of machine components. In AMB spindles, housings support AMBs, backup bearings, air turbine nozzles, and sensors. Housing design and material selection must satisfy the following objectives:
\begin{itemize}
    \item Structural resonance frequencies of the housing must be outside of the operating speed range and dominant disturbance frequencies.
    \item Electrically conducting housing regions exposed to time-varying magnetic fields can incur significant eddy current losses. To minimize these losses, materials with low electrical conductivity (high resistivity) should be used and conductive regions near leakage fields should be avoided.    
    \item Unintended magnetic flux paths must be minimized by using non-magnetic (low permeability) material such as austenitic stainless steels for housing regions in proximity to AMB magnetic circuit.
    \item Thermal conduction from AMB and motor stators must be considered which also relates to the cooling system design.
    \item Sufficient rigidity and support stiffness are required to maintain tight AMB airgaps.
\end{itemize}

\subsection{Step 7: Cooling Design}   \label{sec: Step7:Cooling}
In a typical AMB spindle, the primary sources of temperature rise include Joule (copper) losses in AMB and motor coils, core losses (hysteresis and eddy current) in magnetic materials exposed to time-varying fields, and aerodynamic windage losses on rotating surfaces. A lumped thermal network or FE thermal model is generally used to estimate the temperature rise in critical components to verify that temperature limits are not exceeded. While lumped models are computationally efficient for quick iterations, an FE model allows the study of temperature distribution and reveals coil hot spots. Based on these analyses, appropriate cooling methods, such as forced air cooling, liquid cooling jackets, or enhanced conduction paths, are integrated into the design. Restricting coil current densities to safe ranges for particular modes of cooling also limits temperature rise. A computational fluid dynamics analysis further allows the study of effective convection mechanisms \cite{schweitzerMagneticBearingsTheory2009}. 

In case the cooling performance is not adequate to keep predicted temperatures within safety limits, the design of the cooling system including coolant choice and flow paths should be revisited. This may involve modifying coolant selection, flow paths, or housing conduction characteristics. Where feasible, reducing heat generation at the source, through coil redesign or mitigation of core losses, should be prioritized over increasing cooling capacity.

\subsection{Step 8: Manufacturing and Assembly}   \label{sec: Step8:ManAsm}
The final step involves manufacturing, assembly, and validation of the spindle. This process begins by constructing the spindle CAD assembly to verify that all components sit properly relative to each other. Manufacturing tolerances are specified based on functional requirements, recognizing that ultra-tight tolerances increase cost and should be reserved for surfaces critical to performance, such as AMB air gaps and bearing interfaces.

Possible iterations and adjustments may be necessary in component design based on manufacturing feedback. The procurement of off-the-shelf components takes place side-by-side with manufacturing. These standard parts include fasteners, conventional bearings, magnet wires, sensors, power electronics (inverters and amplifiers), and raw material for manufacturing processes. For critical materials, particularly metals used in high-stress or magnetic components, certified suppliers and inspection documentation are essential to ensure consistency with design assumptions.

The assembly process takes the manufactured and procured components and first constructs sub-assemblies which undergo commissioning steps to verify the final performance of the spindle, for example, short-circuit testing of coils and experimental modal testing of rotor and housing~\cite{kurvinenDesignManufacturingModular2021}. Finally, low-load and full-load tests are performed on the final assembly to verify spindle performance.

\section{Case Study}  \label{sec: CaseStdy}
Based on the proposed multidisciplinary design framework, a high-speed AMB spindle was designed and manufactured for micro-milling applications. The dependence of engineering disciplines on each other and the design flow are thoroughly detailed in this section.

\subsection{Application-specific requirements} \label{sec: slenderrotor app req}

In accordance with Step 1 of the design framework, high-speed micro-milling dynamics and real-time control system require the spindle design to consider application requirements stated in terms of milling forces, unbalance, rotational speed, and negative stiffness. These requirements also influence the material selection and optimization constraints.\\

\subsubsection{Micro-milling forces and unbalance} \label{sec:millforces}

\begin{figure}[h!]
\centering
  \includegraphics[width=0.9\columnwidth]{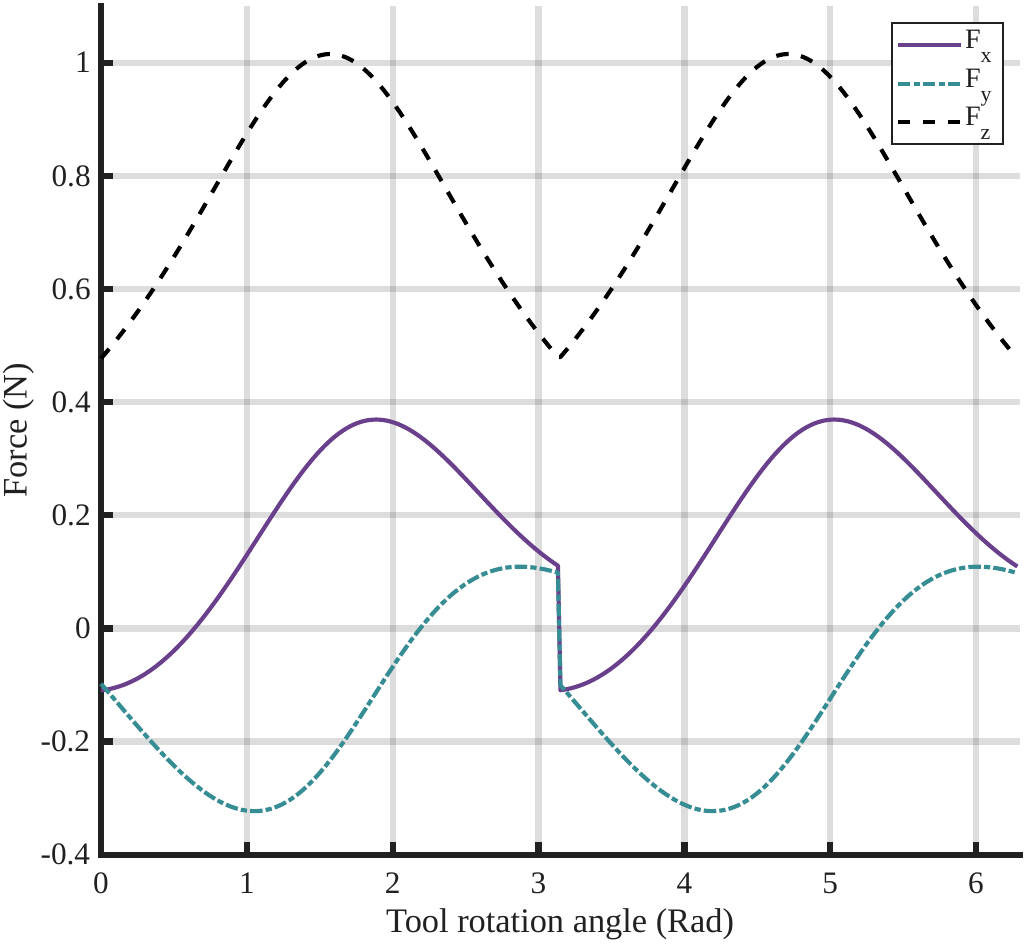}
  \caption{Predicted cutting forces in \textit{x}, \textit{y}, and \textit{z} directions for a slot-milling operation using a two-flute tool.}
  \label{fig:milforces}
\end{figure}

Accurate estimation of machining forces is essential to dimension the AMBs and ensure sufficient load capacity under both static and dynamic operating conditions. In micro-milling, the cutting forces depend strongly on feed per tooth, depth of cut, tool geometry, and workpiece material properties. In this study, the cutting force model proposed by Dow \textit{et al.} \cite{dowToolForceDeflection2004} was employed to estimate the force components acting on the tool and, consequently, on the rotor. The following parameters were selected to represent a typical slot-milling operation, with the axis of the spindle aligned along \textit{z} direction: two-flute tool diameter (0.2 mm), up-feed (5 $\mathrm{\mu m/flute}$), depth of cut (10 $\mathrm{\mu m}$), and workpiece (S-7 tool steel with hardness 5.5 GPa). Figure \ref{fig:milforces} shows the predicted cutting forces in \textit{x}, \textit{y}, and \textit{z} directions with static (mean) components of 0.17, -0.12, and 0.77 N, respectively. 

In addition to machining forces, the permissible rotor unbalance force \(F_u\) in N for a rigid rotor was calculated using the International Standard ISO-1940-1:2003 as
\begin{equation}
    F_{u} = \frac{G m n}{1000}   
\end{equation}
where \(G\) is the balance quality grade in mm/s, \(m\) is the rotor mass in kg, and \(n\) is the rotor speed in rad/s. With sufficiently tight balance tolerance, we aim for $G = 1$. With the mass of rotor (described in section~\ref{sec: rotorDes}) as 0.22~kg and rotational speed of \(1.15\times10^4\) rad/s (110,000 rpm as chosen in section~\ref{sec:rotspeed}), we determined \(F_u\) = 2.54~N. \\

\subsubsection{Rotational speed} \label{sec:rotspeed}

Compared to macro-scale milling, the smaller diameter cutting tools (as small as 50~$\mu$m) used in micro-milling result in reduced cutting speeds and hence low material removal rates. To attain acceptable process efficiency (\emph{i.e.}, to attain high material removal rates), cutting tools in micro-milling process must rotate at substantially higher rotational speeds ~\cite{chaeInvestigationMicrocuttingOperations2006}. In many applications, miniature micro-machining spindles commonly operate above 60,000 rpm~\cite{ehmannInternationalAssessmentResearch2005, bedizDynamicsUltrahighspeedUHS2014}.

To determine the required rotational speed for the present design, a minimum micro tool diameter of 200~$\mu$m was initially considered; assuming a typical cutting speed of 130 m/min, which is sufficient to machine various common materials. This yields a rotational speed of 206,901 rpm in theory. However, the design of the AMB-supported rotor involves a trade-off between rotational speed and axial AMB load capacity. Higher speeds require a reduction in rotor thrust disc diameter to maintain safe centrifugal stress levels. In contrast, axial AMB requires sufficient disc surface area to generate an adequate thrust force. Based on initial feasibility analysis, the target rotational speed was reduced to increase the maximum allowed disc diameter (and hence pole surface area) for sufficient axial AMB load capacity. Therefore, we settled on a lower rotational speed of 110,000 rpm (1833.33 Hz) while also reviewing the target speeds of other AMB spindles~\cite{kimmanMiniatureMillingSpindle2010,parkMagneticallySuspendedMiniature2012}. At this rotational speed, the cutting speed is 69.1~m/min with a 200~$\mu$m tool, while the cutting speed of 130~m/min is achievable with a 380~$\mu$m tool. Acknowledging that rotor design can accommodate a range of tool diameters, the selected rotational speed represents a balanced compromise between machining performance, structural integrity, and axial AMB load capacity.\\

\subsubsection{Negative stiffness} \label{sec:negstiff}

As a guideline, AMB needs to be dimensioned to create a negative stiffness of the same order, or at most one order lower, as the desired controlled (positive) stiffness~\cite{molenaarNovelPlanarMagnetic2000}. The influence of disturbances such as cutting forces on the rotor needs to be analyzed to determine the required controlled stiffness needed to compensate for static cutting forces. This can be achieved by considering AMB as a spring-mass system (where mass refers to the mass of the rotor; stiffness refers to the spring stiffness required to compensate for the disturbances) followed by plotting the power spectral density of the cutting force disturbance. As detailed in \cite{kimmanDesignMicroMilling2010}, a closed-loop controlled stiffness of \(1.4\times10^5\) N/m is determined.

\subsection{Rotational Drive} \label{sec: airturbine}
High-speed spindles are commonly driven by electric motors or air (pneumatic) turbines. For instance, Park \textit{et al.} \cite{parkMagneticallySuspendedMiniature2012} utilized an air turbine, while Kimman \textit{et al.} \cite{kimmanMiniatureMillingSpindle2010} used a permanent magnet synchronous motor. In the present study, an air turbine was selected following the design methodology of Li \textit{et al.} \cite{liHighspeedPrecisionMicrospindle2019,liDesignEvaluationHighspeed2015} due to its structural simplicity, reduced thermal load, and inherent suitability for ultra-high-rotational speeds. In this configuration, compressed air is supplied through nozzles and directed onto turbine buckets machined directly into the rotor.\\

\subsubsection{Aerodynamic Design}

As a first design step, turbine pitch diameter \(d_m\) was verified to keep the Mach number \(Ma\) subsonic to avoid shock-related compressibility effects. Since the Pelton-type impulse turbine buckets were to be directly machined on the rotor, \(d_m\) was constrained to be slightly less than the suitable rotor segment of 12~mm diameter. The ratio of air flow velocity \(u\) to sound velocity \(c\) defines \(Ma\) as
\begin{equation}
Ma = \frac{u}{c} = \frac{\pi n_{t} d_m }{60 c}    
\end{equation}
where \(n_{t}\) is turbine rpm. With the target rotational speed of spindle as \(n_{t}\) = 110,000 rpm and \(d_m\) = 10~mm, the resulting \(Ma\) is 0.17, which is well below the typical onset of transonic effects at \(Ma = 0.8\).

The second design step ensured that the available supply pressure from air compressor could provide the required rotational speed to the turbine. Considering an isentropic expansion in nozzle from inlet pressure \(P_{in}\) to outlet pressure (backpressure) \(P_{out}\) = 0.1 MPa, the nozzle exit velocity is~\cite{liHighspeedPrecisionMicrospindle2019}
\begin{equation}
    u = \sqrt{\frac{2 \kappa}{(\kappa-1)} R T_o\left[1-\left(\frac{P_{out}}{P_{in}}\right)^{\frac{\kappa-1}{\kappa}}\right]}
\end{equation}
where \(\kappa\) = 1.4 is the specific heat ratio for dry air, \(R = 287.05\)~J/(kg K) is the specific gas constant for dry air, and \(T_o\) is the absolute temperature of air source. We assumed \(T_o = 293.15\) K corresponding to a compressor tank that has cooled to ambient. Next, \(u\) can be related to ideal turbine speed in rpm \(N\) while accounting for airflow loss coefficient \(\phi\) and turbine speed efficiency \(\xi\) as
\begin{equation}
    N = \frac{60 u \phi \xi}{\pi d_m}
\end{equation}
where typical range for \(\phi\) is 0.92-0.98~\cite{liHighspeedPrecisionMicrospindle2019}. Opting for an ultra-conservative design with \(\phi\) = 0.92 and \(\xi\) = 0.2, \(P_{in}\) of 0.2~MPa yields \(N\) = 114,320~rpm, exceeding the required rotational speed of 110,000~rpm. Note that steady turbine speed under machining loads will be lower than \(N\) for the same supply pressure because turbine torque must balance machining and aerodynamic torques. In other words, a higher supply pressure may be needed to sustain the required turbine speed during machining. The selected air compressor (Einhell Silent TE-AC 24) is capable of delivering up to 0.9~MPa absolute pressure, providing sufficient margin.

\begin{figure}[h!]
\centering
  \includegraphics[width=0.9\columnwidth]{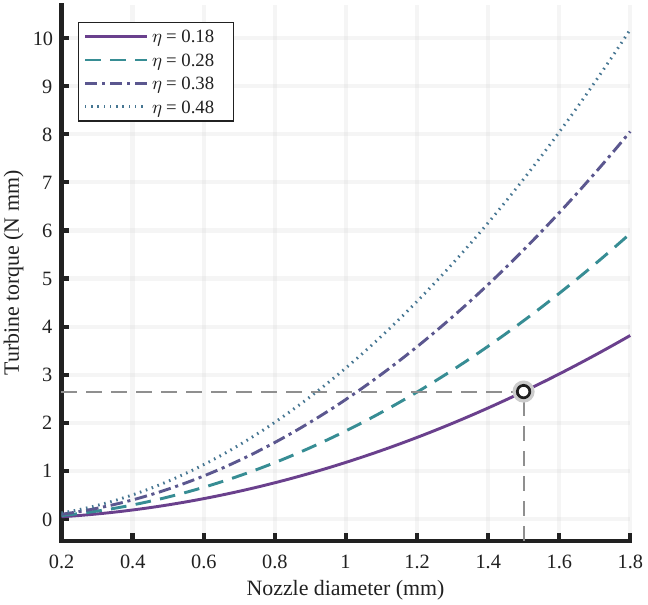}
  \caption{Turbine torque as a function of nozzle diameter with turbine power efficiency range 0.18-0.48. Selected nozzle diameter 1.5 mm results in 2.65 N mm of torque with conservative power efficiency of 0.18.}
  \label{fig:Torque}
\end{figure}

The third design step involved the selection of a nozzle diameter such that the available turbine torque at the operating point exceeded the load torques (windage and machining). The jet power from nozzles \(P_{jet}\) and turbine shaft power \(P_{shaft}\) are given as
\begin{equation}
    P_{jet} = \frac{1}{2}ZGu^2\;\;\;\;\;\;\;\;\;\;\;\;\;P_{shaft} = \frac{\pi M N}{30} 
\end{equation}
where \(Z\) is the number of nozzles, \(G\) is the air mass flowrate, and \(M\) is the turbine output torque. The mass flowrate is given by
\begin{equation}
    G = \rho_{out} u A_n = \rho_{out} u \frac{\pi d_n^2}{4}  \label{Eqn: G}
\end{equation}
where \(\rho_{out}\) is the air density at nozzle exit, \(A_n\) is the nozzle area, and \(d_n\) is the nozzle diameter. Using the isentropic density relation and ideal gas law, \(\rho_{out}\) can be written as
\begin{equation}
    \rho_{out} = \frac{P_{in}}{R T_{in}} \left( \frac{P_{out}}{P_{in}} \right)^{\frac{1}{k}}  \label{Eqn: rho_out}
\end{equation}
The power equations above are related by power efficiency \(\eta\) as: \(P_{shaft} = \eta P_{jet}\). After all relevant substitutions in this power balance, the turbine output torque is given as
\begin{equation}
    M = \frac{\pi \eta Z d_m d_n^2 P_{in} u^2}{16 \phi \xi R T_{in}} \left( \frac{P_{out}}{P_{in}} \right)^{\frac{1}{k}}
\end{equation}
where \(T_{in}\) is the temperature of air source taken as 293.15~K. With \(Z\)= 4 and other parameters as mentioned in the discussion above, Figure~\ref{fig:Torque} plots the variation of turbine output torque with nozzle diameter for a range of power efficiencies 0.18-0.48.

Next, the load torques on the rotor from machining and windage were estimated. The machining torque was determined by evaluating tangential forces on cutting tool in a slot milling operation using the force model by Dow \textit{et al.} \cite{dowToolForceDeflection2004}. For a conservative design, more demanding machining conditions were selected compared to those in section~\ref{sec:millforces}. With a tool diameter of 1~mm, an up-feed of 10~\(\mu\)m/flute, depth of cut of 10~\(\mu\)m, and S-7 tool steel with hardness 5.5~GPa, the maximum tangential force on the cutting tool is 1.93~N which resulted in a machining torque of 0.97~N mm. The windage torque on rotor cylindrical surfaces facing radial AMB was determined using the friction coefficients from Bilgin and Boulos \cite{bilgenFunctionalDependenceTorque1973}. With a windage torque of 0.18~N mm, the total load torque (machining and windage) is 1.15~N mm. From Figure~\ref{fig:Torque}, with a conservative power efficiency of 0.18, a nozzle diameter of 1.5~mm was selected to give 2.65~N mm of turbine output torque which comfortably exceeded the total load torque.

\begin{figure}[h!]
\centering
  \includegraphics[width=0.98\columnwidth]{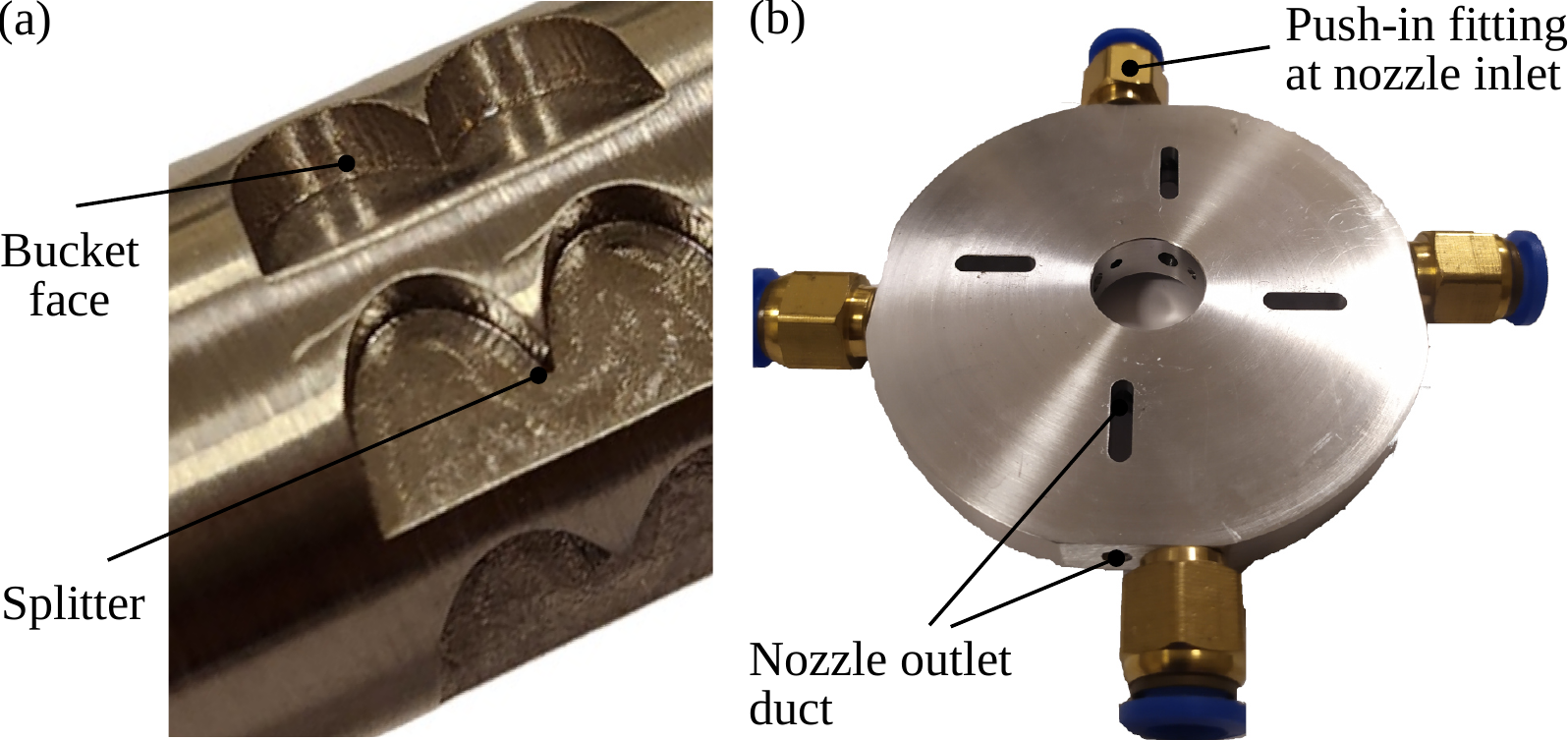}
  \caption{Air turbine components. (a) Turbine buckets and splitter machined into the rotor. (b) Nozzle plate with push-in pneumatic fittings.}
  \label{fig:AirTurbine}
\end{figure}

Pelton-inspired impulse turbine buckets were designed and machined directly into the rotor. As shown in Figure~\ref{fig:AirTurbine}(a), a central splitter divides the air jet into two buckets. As with Pelton turbines, this division in the ideal symmetric case reduces axial thrust and hence the disturbance load on axial AMB. The buckets provide the air flow with a continuous curvature to avoid flow separation. 

A higher bucket number generally reduces torque ripple by increasing overlap of jet-bucket engagement cycles. Within the constraint of the rotor diameter of 12 mm, we maximized the bucket count to 7. This number was also chosen intentionally to avoid it being an integer multiple of the nozzle count (\(Z=4\)). This arrangement prevents the simultaneous impact of all four jets on the buckets for reduced amplitude of torque and force fluctuations. The bucket topology was inspired by the reference design approaches described by Shi and Dong~\cite{shiSyntheticOptimizationAir2014}, Solemslie and Dahlhaug~\cite{solemslieReferencePeltonTurbine2014}, and Juraeva \textit{et al.} \cite{juraevaDesigningHighspeedDental2017}. While the current profile serves as a functional baseline, optimization using computational fluid dynamics is left for future work. Major parameters of the air turbine are shown in Table~\ref{tab:AirTurbine}.

\begin{table}[h]
    \caption{Air turbine parameters.}
    \label{tab:AirTurbine}
    \centering
    \begin{tabular}{l|l|l}
        \hline
        Pitch diameter          & \(d_m\)       & 10 mm           \\ 
        Nozzle diameter         & \(d_n\)       & 1.5 mm           \\ 
        Number of nozzles        & \(Z\)       & 4           \\ 
        Number of buckets        & \(Z_{b}\)       & 7           \\ 
        Inlet pressure        & \(P_{in}\)       & 0.2 MPa           \\ 
        Design operating speed    & \(N\)             & 114,320 rpm           \\ 
        Turbine output torque    & \(M\)       & 2.65 N mm           \\
        \hline
    \end{tabular}
\end{table}

\subsubsection{Realization and Hardware Procurement}
Air turbine buckets were machined into the rotor using standard CNC milling operations. Based on the pressure requirements, an Einhell Silent TE-AC 24 air compressor which has a 24 L tank capacity, maximum operating pressure of 8 bar (gauge) (0.9~MPa absolute), and low noise level of 57~dB(A), was selected. Using polyurethane pipes, the compressor connects to a 1-to-4 pneumatic port which supplies air to 4 nozzles of the air turbine. The nozzle plate is shown in Figure~\ref{fig:AirTurbine}(b). Push-in pneumatic fittings were used for connections as these are typically rated up to 1~MPa of pressure. Outlet ducts were drilled into the nozzle plate with open slots to allow some air to escape axially outward within the spindle providing a limited cooling effect on the coils.

\subsection{Rotor}   \label{sec: rotorDes}

To support the functionalities of radial and axial AMBs, displacement sensors, rotational drive, and milling tool, the rotor is divided into several segments. The design and manufacturing of the rotor are discussed in this section.\\

\subsubsection{Mechanical Design}

For AMB applications, the rotor material should exhibit high magnetic permeability and low coercivity to minimize hysteresis losses and maximize force capacity while high electrical resistivity and thin laminations are needed to minimize eddy current losses. Electrical steel laminations are commonly employed in rotor segments facing the poles of radial AMB due to their favorable magnetic properties. However, at ultra-high rotational speeds, centrifugal forces may compromise the mechanical integrity and tight fit of laminated stacks. To eliminate this risk, a solid rotor configuration was adopted in the present design. Suitable materials for solid rotor are those that exhibit soft magnetic properties with an intrinsic coercivity below 1000 A/m. Ferritic stainless steels satisfy this criterion but generally exhibit limited mechanical strength. To ensure structural robustness at high speeds, martensitic stainless steel AISI~410 was selected. Although its magnetic properties are inferior to those of ferritic grades, they can be improved through annealing. 

\begin{figure}[h!]
\centering
  \includegraphics[width=0.9\columnwidth]{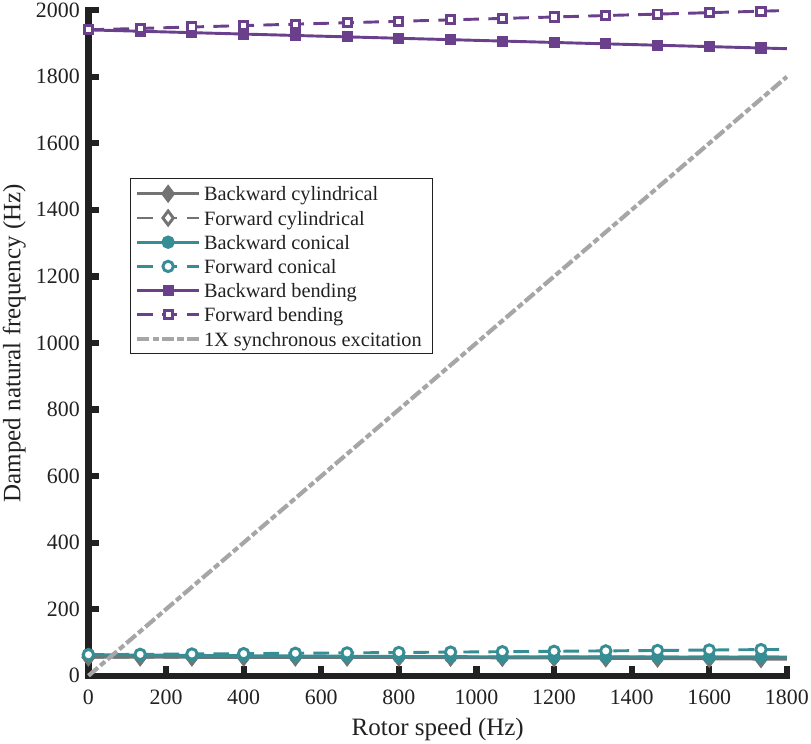}
  \caption{Rotor Campbell diagram showing rigid body (cylindrical and conical) and bending modes.}
  \label{fig:spindlecampbell}
\end{figure}

After material selection, the diameter of a high-speed rotor needs to be dimensioned to keep centrifugal stress within safe limits. The maximum attainable rotational speed $\Omega_\mathrm{max}$ can be related to its radius \textit{r} as
\begin{equation}
    \Omega_\mathrm{max} = \frac{1}{r} \sqrt{\frac{8\sigma_o}{(3+\nu)\rho}}
    \label{eqn: rotorspeed}
\end{equation}
where $\sigma_o$ is the yield strength, $\nu$ is the Poisson's ratio, and $\rho$ is the density of the material \cite{larsonneurDesignControlActive1990}. With rotational speed requirement of $\Omega_\mathrm{max}$ = 110,000 rpm and AISI 410 material properties of $\sigma_\mathrm{o}$~=~275~MPa, $\nu$~=~0.28, and $\rho$~=~7700~kg/m$^{3}$, the theoretical maximum diameter is 51.2 mm. Considering material inhomogeneities, a conservative factor of safety was applied, and we set the rotor disc diameter at 30 mm. The rotor segments in radial AMBs have a diameter of 16.4 mm to meet the target diameter requirement of the displacement sensor (PRI04, Sensonics, UK) and to accommodate typical micro-machining tool shank diameters. The rotor segment for the rotational drive (air turbine) is kept at 12 mm in diameter as discussed in section \ref{sec: airturbine}. The overall rotor length is kept at 155 mm, accommodating other spindle components and maintaining adequate separation between aimed closed-loop control bandwidth (300~Hz) and first bending (flexural) mode of rotor. The rotor design also kept the first bending mode out of the operating range (0-110,000 rpm or 0-1833.33 Hz) as seen in the rotor Campbell diagram in Figure~\ref{fig:spindlecampbell} computed using the open-source rotordynamic code VibronRotor~\cite{ahmedVibronRotorOpensourceRotordynamic2019}.

\subsubsection{Manufacturing}
\begin{figure}[h!]
\centering
  \includegraphics[width=0.9\columnwidth]{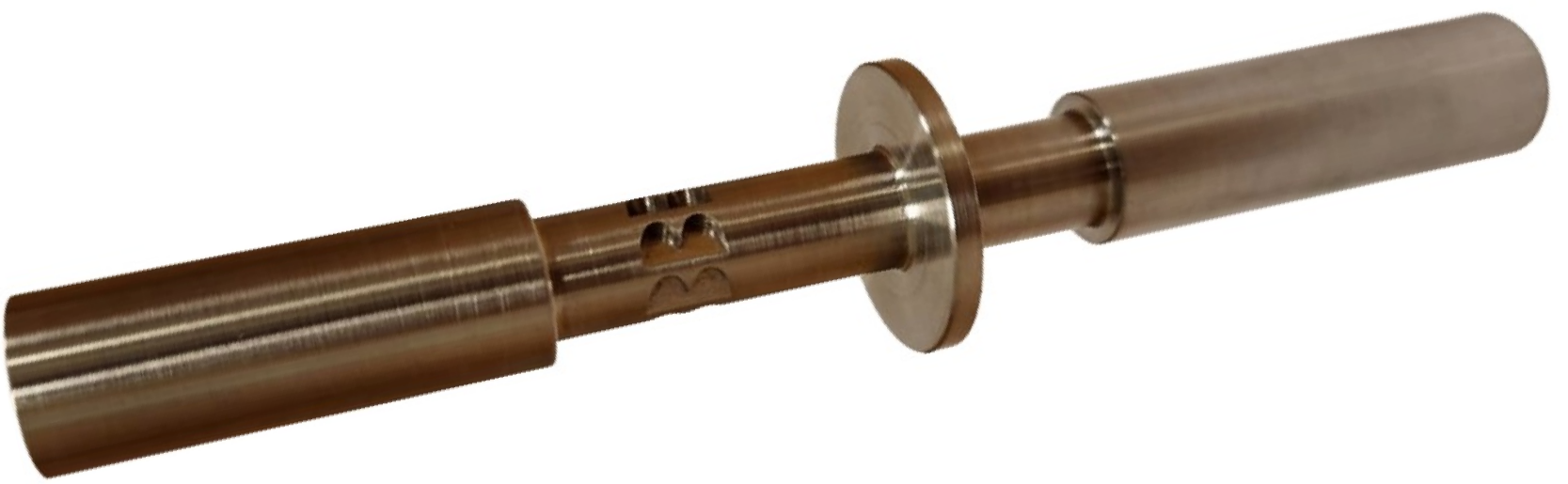}
  \caption{Machined rotor.}
  \label{fig:longrotor}
\end{figure}

The rotor was machined from AISI 410 bar stock (Ugitech SA, France) and subsequently finished by cylindrical grinding to achieve high roundness and low surface roughness. This minimizes synchronous runout in displacement sensor signals and hence leads to enhanced control performance. In particular, high (superior) cylindricity prevents the control system from interpreting geometric imperfections as physical rotor displacements. For improved magnetic softness, annealing is generally recommended but it also degrades the mechanical strength of the material. Heat treatment can introduce distortions on the rotor surface, requiring re-grinding. Ideally, the part should be annealed before the finishing process of grinding. Afterwards, in accordance with ISO-1940-1:2003, the rotor will be dynamically balanced with an aimed quality grade of 1. The manufactured rotor is shown in Figure~\ref{fig:longrotor}.

\subsection{Radial AMB} \label{sec: radialAMB}

In accordance with Step 4 of the proposed design framework, a comprehensive AMB design including analytical modeling, FE validation, multi-objective optimization, and hardware realization is presented here. Each radial AMB (see Figure~\ref{fig:RadialAMBCS}) controls the rotor in two translational degrees of freedom (DOFs), namely \textit{x} and \textit{y}. Two identical units are mounted at opposite ends of the rotor as shown in Figure~\ref{fig:spindlelabelled}, which illustrates the assembly of the system, providing  4 (radial) DOFs in total.\\

\subsubsection{Topology and modeling}   \label{sec: Topology}

\begin{figure}[h!]
    \centering
  \includegraphics[width=0.9\columnwidth]{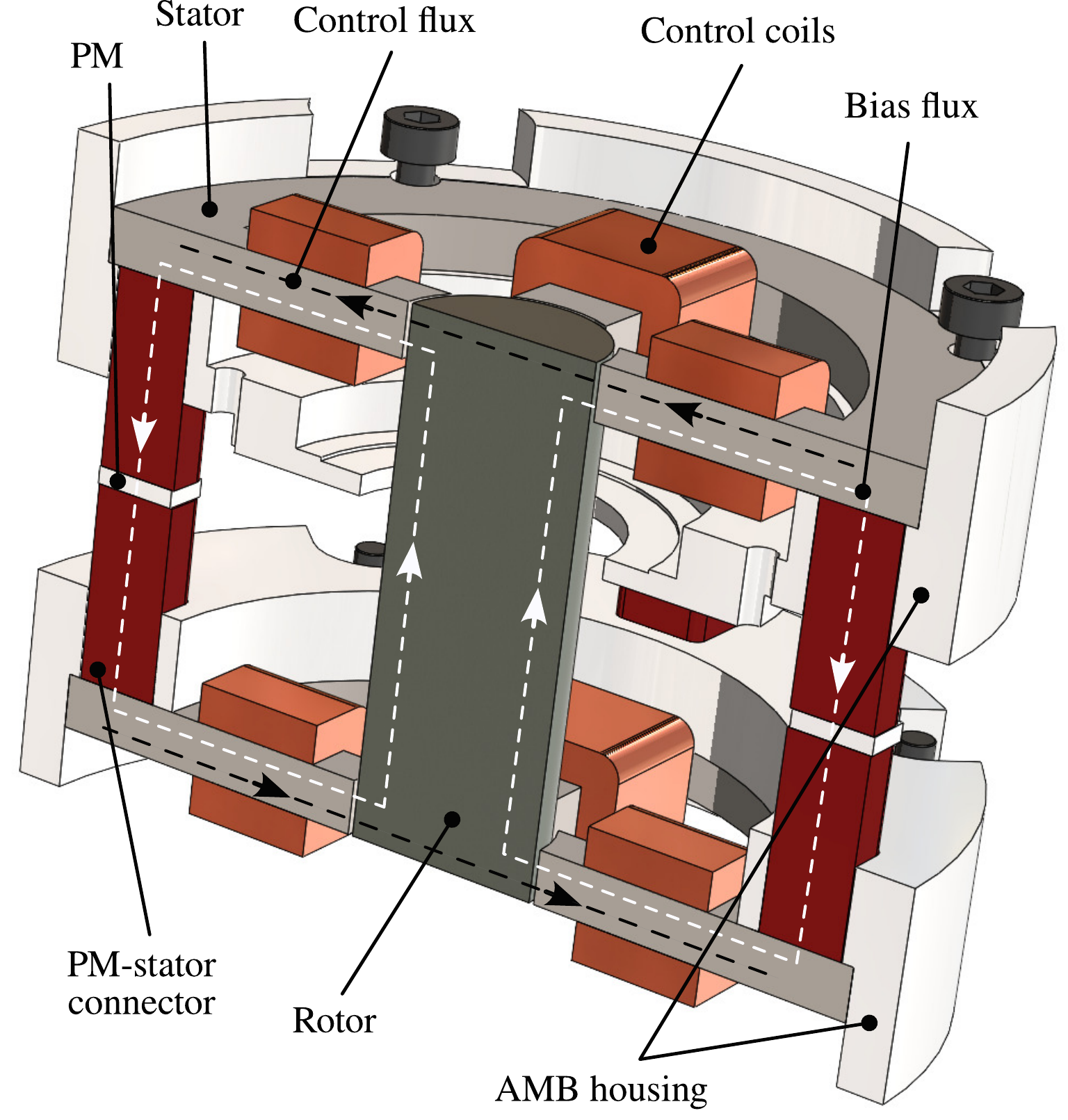}
  \caption{Radial AMB cross-section with bias and control flux paths.}
  \label{fig:RadialAMBCS}
\end{figure}

A permanent magnet (PM)-biased homopolar AMB topology is adopted in this study. Homopolar AMB design maintains a uniform pole polarity in a given rotational plane, reducing hysteresis and eddy current losses in the rotor, compared to heteropolar designs. The use of a solid rotor as mentioned in section \ref{sec: rotorDes} further supports the need for a homopolar topology. As shown in Figure~\ref{fig:RadialAMBCS}, this configuration consists of two four-pole stators connected to four PMs using PM-stator connectors. PMs generate a bias flux, while control coils wound around each pole of the stators produce control flux. The control flux remains within the stators and interacts with bias flux to create a difference in magnetic flux densities in opposite radial airgaps to produce radial forces. This way, AMB generates the forces as intended by the control system. 

In materials, M270-35A electrical steel laminations were selected for stators, Neodymium magnets (NdFeB N35) for PMs, and annealed AISI 430F ferritic stainless steel for PM-stator connectors. Since forced cooling was not included in this design, the coil current density was kept at around 4~A/mm\(^2\), consistent with recommended values of 2-5~A/mm\(^2\) for naturally cooled magnetic bearings and electrical machines~\cite{khooSpecificLoadCapacity2007}.

The equivalent magnetic circuit model forms the analytical model for the radial forces in this AMB. Under linear material assumptions, the flux leakage and magnetic saturation effects are ignored, and an infinite core permeability is assumed. Using Maxwell's equations, bias and control flux densities in radial airgap are derived and added due to the superposition principle \cite{kimmanDesignMicroMilling2010} to determine the radial bearing force on the rotor. This force depends non-linearly on control current and airgap length, hence it is linearized at its equilibrium position (rotor centered and zero control current) for linear control design \cite{molenaarNovelPlanarMagnetic2000} using first-order Taylor expansion. The linearized radial force is given as
\begin{equation}
    F_x = K_x\;x + K_{ix}\;i_x 
\end{equation}
where $K_x$ is the radial force-displacement dependency (also called radial negative stiffness), $x$ is the rotor radial displacement, $K_{ix}$ is the force-current dependency, and $i_x$ is the control current in radial coils. The dependencies can be written as follows
\begin{align}
K_{x} &= \frac{8 \mu_{o} k_m A_{gr} A_{m}^{3} B_{r}^{3} H_{c}^{2} l_{m}^{2} }
{\Big( 2 k_m A_{m} B_{r} l_{gr} + \mu_{o} k_l A_{gr} H_{c} l_{m} \Big)^{3}} \\
K_{ix} &= \frac{4 \mu_0 A_{gr} A_{m} B_{r} H_{c} l_{m} n_{r} } 
{k_{c} l_{gr} \Big( 2 k_m A_{m} B_{r} l_{gr} + \mu_0 k_l A_{gr} H_{c} l_{m} \Big)} \label{eqn:Ki}
\end{align}
where \(\mu_{0}\) is the magnetic permeability of air, \(A_{gr}\) is the radial stator pole area, \(A_{m}\) is the PM area, \(B_r\) is the PM remanent flux density, \(H_{c}\) is the PM coercive force, \(l_{m}\) is the PM thickness, \(l_{gr}\) is the radial airgap, \(n_{r}\) is the radial coil turns per pole. The correction factors \(k_{m}\) and \(k_{c}\) account for the assumption of infinite permeability in bias flux and control flux circuits, respectively, while \(k_l\) deals with the assumption of no flux leakage. An electromagnetic three-dimensional (3D) FE model, was developed to determine these correction factors.

The electromagnetic 3D FE model takes into account the effects neglected in the analytical model such as flux leakage, finite permeability of materials, and magnetic saturation. The comparison of airgap flux densities from analytical and FE models determines the correction factors. We utilized the magnetic fields (mf) interface of the AC/DC module of \textit{COMSOL Multiphysics} software to construct the FE model. This interface solves Maxwell's equations formulated based on magnetic vector potential and scalar electric potential as dependent variables. For AMB components, Ampere's Law node was used that defines magnetization models such as magnetic flux density-magnetic field intensity (referred to as B-H) curves and relative permeability. For control coils, coil node with homogenized multi-turn conductor model was used. All AMB components were enclosed in an air region so that flux leakage and fringing effects became part of the analysis. Stationary relative tolerance was set at \(1\times10^{-4}\). The Coil Geometry Analysis and main Stationary steps were run with direct solvers MUMPS (multifrontal massively parallel sparse direct solver) and PARDISO (parallel sparse direct solver), respectively. We successively refined the mesh to perform a mesh-independence study based on grid convergence index (GCI) \cite{roachePerspectiveMethodUniform1994, celikProcedureEstimationReporting2008}. With AMB forces (in \(x\) and \(y\) directions) showing non-monotonic behavior on refinement, we determined a GCI band based on three plausible orders of convergence $(p = 1,1.5,2)$ to verify the asymptotic behavior of 3D mesh (tetrahedral) refinement. Comparing finer and extra-fine physics-defined meshes (refinement ratio of 1.309), GCI on extra-fine mesh ranges between 0.24\textendash0.54\% for $x$-forces and 0.03\textendash0.07\% for $y$-forces, providing an acceptable discretization uncertainty bound. The tetrahedral elements were limited to minimum and maximum sizes of 0.116 and 2.7 mm, respectively. The model was composed of 691,368 DOFs with a solution time of 6 min 15 s on an eight-core Intel i7-11850H processor and 16 GB RAM. Peak physical and virtual memory usage were 8.91 GB and 20.28 GB, respectively.\\

\subsubsection{Optimization model}

The optimization model constructed here optimizes the AMB performance in relation to micro-milling process requirements as outlined in section~\ref{sec: slenderrotor app req}. One important AMB performance metric to maximize is the specific load capacity which is the ratio of the maximum electromagnetic force the AMB can generate (i.e. the load capacity) and its volume. A lower volume, and hence a lower mass, aids the goal of spindle miniaturization and permits the use of cost-effective linear stages to undergo accelerations required by machining processes. Therefore, a multi-objective optimization problem was formulated to maximize static radial load capacity while minimizing AMB volume, to improve specific load capacity and enabling spindle miniaturization. The design variables, shown in Table~\ref{tab:DesVar}, include the geometrical dimensions of AMB stator and PM as well as the control current. An important requirement for linear control design is to operate the AMB within the linear region of the material B-H curves. For the stator and rotor materials used in this study, the curves are almost linear up to 1~T. Thus, we limited the total airgap flux density to 1~T. To prevent magnetic field reversal in the differential driving AMB mode, we restricted the magnitude of control flux density to be less than the bias flux density in the airgap. Further, the negative stiffness was limited between \(1.4\times10^4\)\textendash\(1.4\times10^5\)~N/m as detailed in section~\ref{sec:negstiff}.

\begin{table}[h]
    \caption{Radial AMB design variables at selected design point.}
    \label{tab:DesVar}
    \centering
    \begin{tabular}{l|l|l}
        \hline
        Stator inner radius        & $s_{ir}$      & 27.1 mm                       \\ 
        Stator radial thickness    & $s_{rt}$      & 9.1 mm                      \\ 
        Pole length                & $p_l$         & 8.4 mm                      \\ 
        Pole width                 & $p_w$         & 4.7 mm                       \\ 
        Pole radius                & $p_r$         & 8.6 mm                       \\ 
        PM thickness               & $pm_t$        & 1.5 mm                      \\     
        PM width                   & $pm_w$        & 8 mm                      \\ 
        PM length                  & $pm_l$        & 7 mm                      \\ 
        Control current            & $i_c$         & 2.98 A                      \\ 
        \hline
    \end{tabular}
\end{table}

Since heuristic methods such as the genetic algorithm (GA) possess the ability to converge to a global minimum, we utilized GA in MATLAB (\textit{gamultiobj}) for our multi-objective optimization problem to develop a Pareto front. Such a Pareto front provides the designer with freedom to select design points for analysis and production that adhere to different application requirements. The optimization algorithm terminated based on its spread-based stopping criteria (average change in Pareto solutions spread fell below \textit{FunctionTolerance} of \(1\times10^{-4}\)) and returned a non-dominating set approximating a Pareto front in 18.5 seconds on an eight-core Intel i7-11850H processor and 16 GB RAM. The options set used for MATLAB's \textit{gamultiobj} function are listed in Table~\ref{tab:GAsettings}.

\begin{table}[h]
    \caption{Optimization algorithm (GA) settings.}
    \label{tab:GAsettings}
    \centering
    \begin{tabular}{l|l}
        \hline
        Maximum generations        & 1800                            \\ 
        Population size            & 200                             \\ 
        Maximum stall generations  & 100                             \\ 
        Distance measure function  & Distance crowding (phenotype)    \\ 
        Pareto fraction            & 0.35                             \\ \hline
    \end{tabular}
\end{table}

\subsubsection{Analysis} \label{sec: Radial-AMB-Analysis}

\begin{figure}[h!]
\centering
  \includegraphics[width=0.98\columnwidth]{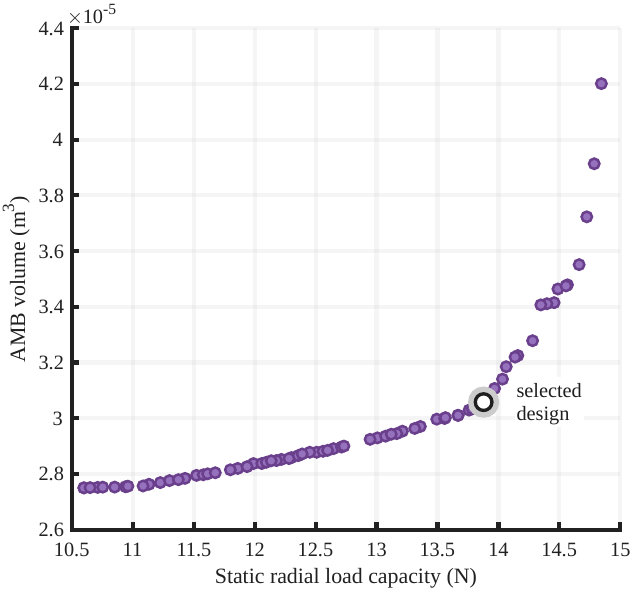}
  \caption{Pareto front with maximization of static radial load capacity and minimization of AMB volume. A selected design point (13.88 N and 3.06$\times$10\(^{-5}\) m\(^3\)) is also shown for further analysis and manufacturing.}
  \label{fig:Pareto}
\end{figure}

The resulting Pareto front is depicted in Figure~\ref{fig:Pareto} with objective 1 (static radial load capacity) on the $x$-axis and objective 2 (AMB volume) on the $y$-axis. These design points cover a range of objectives allowing the designer to choose a suitable design for further analysis and manufacturing. We selected a design point where static load capacity is 13.88 N and AMB volume is \(3.06\times10^{-5}\) m\(^3\). Here, we developed force/frequency curves to ensure that AMB load capacities exceed the load demands. We also analyzed the selected design point with a 3D FE model to verify the chosen correction factors.  

The dynamic load capacity \(F_{d}(\omega)\) is limited by the maximum slew rate of AMB and is a function of disturbance excitation frequency \(\omega\). It was evaluated as
\begin{equation}
    F_{d}(\omega) = \frac{K_{ix}V_{c}}{L\omega}
\end{equation}
where \(K_{ix}\) is the force-current dependency (refer to Eq.~\ref{eqn:Ki}), \(V_{c}\) is the voltage applied by the amplifier to coils, and \(L\) is the coil inductance. The capacity envelope is defined by the minimum of both the static and dynamic capacities, and their intersection is called the knee point. Next, we evaluated the load demands from machining and unbalance forces to confirm that AMB capacity envelope exceeds them by an appropriate margin. From the machining forces predicted in section~\ref{sec:millforces}, static radial force demand (resultant of \textit{x} and \textit{y} mean forces) was determined as 0.213~N. Dynamic force demand was estimated by treating the in-plane milling forces (\textit{x} and \textit{y} components) as a periodic signal over one tool revolution and expressing them in terms of complex Fourier series. The harmonic magnitudes in this series were taken as dynamic components. For a two-flute tool, the peak magnitude at the tooth-passing frequency (2$\times$ rotor frequency) was noted as machining dynamic force demand at 0.214~N. The dynamic load demand from rotor unbalance was estimated in section~\ref{sec:millforces} as 2.54~N synchronous to 1$\times$ rotor frequency. Afterwards, safety factors were applied to each load demand to account for uncertainty in the estimated forces and to retain sufficient margins to the available AMB load capacities~\cite{swansonNEWACTIVEMAGNETIC2014}: 1.5 to the unbalance demand (well-bounded by specified balance grade), 2 to the machining static demand, and 3 to the machining dynamic demand (greater sensitivity to force model inputs and process condition variability).

\begin{figure}[h!]
\centering
  \includegraphics[width=0.98\columnwidth]{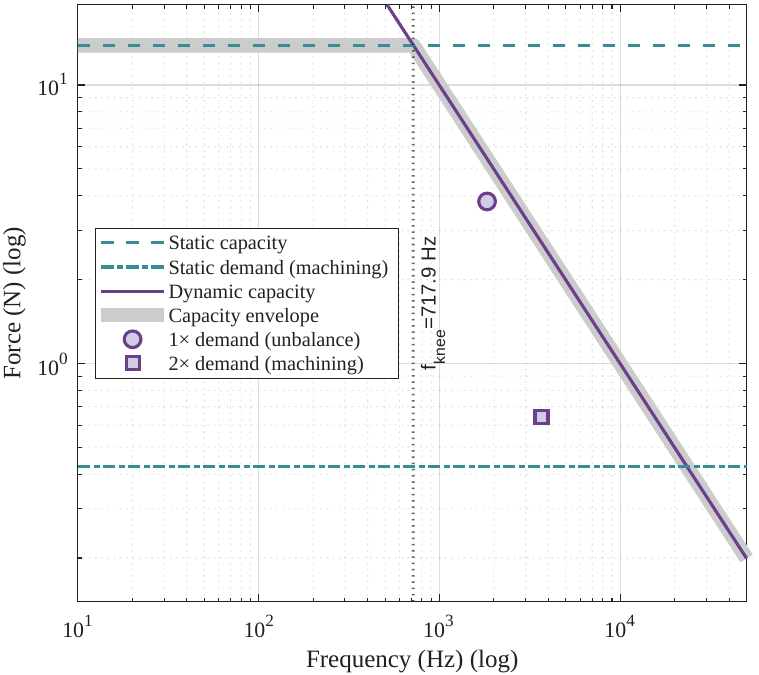}
  \caption{AMB load capacities and demands. The AMB capacity envelope is defined by the minimum of both static and dynamic capacities and is shown with a grey thick line. The static and dynamic loads shown are contained within this capacity envelope for proper AMB operation.}
  \label{fig:AMBForces}
\end{figure}

The static and dynamic AMB load capacities and demands have been plotted in Figure~\ref{fig:AMBForces} after applying the afore-mentioned safety margins to the load demands. The static load capacity of 13.88~N is significantly higher than machining static demand of 0.43 N. The dynamic load demand from rotor unbalance stands at 3.81~N with 1$\times$ rotor frequency while dynamic capacity is 5.43~N at this frequency. Next, machining dynamic demand is 0.64~N at 2$\times$ rotor frequency while dynamic capacity at this frequency is 2.72~N. Hence, all capacities exceed the demands to a good extent. The knee frequency was identified as 717.9~Hz which is the intersection of static and dynamic load capacities. At frequencies higher than the knee frequency, the AMB capacity starts declining approximately proportional to \(1/\omega\). Although this analysis builds confidence in the ability of selected AMB design to meet load requirements, the dynamic load demands at 1$\times$ (rotor frequency) and 2$\times$ (tooth-passing) frequencies occur well beyond the achievable closed-loop bandwidth of AMB, which is limited to only a few hundred Hz as seen in comparable AMB spindles \cite{kimmanMiniatureMillingSpindle2010}. Therefore, AMB will not actively compensate these high-frequency dynamic forces. However, the force versus frequency plots in Figure~\ref{fig:AMBForces} verify that dynamic demands stay within the load capacity envelope, ensuring that AMB hardware does not exceed its dynamic (slew-rate limited) capacity at these high frequencies and will not drive the amplifier into saturation.

\begin{figure}[h!]
\centering
  \includegraphics[width=0.98\columnwidth]{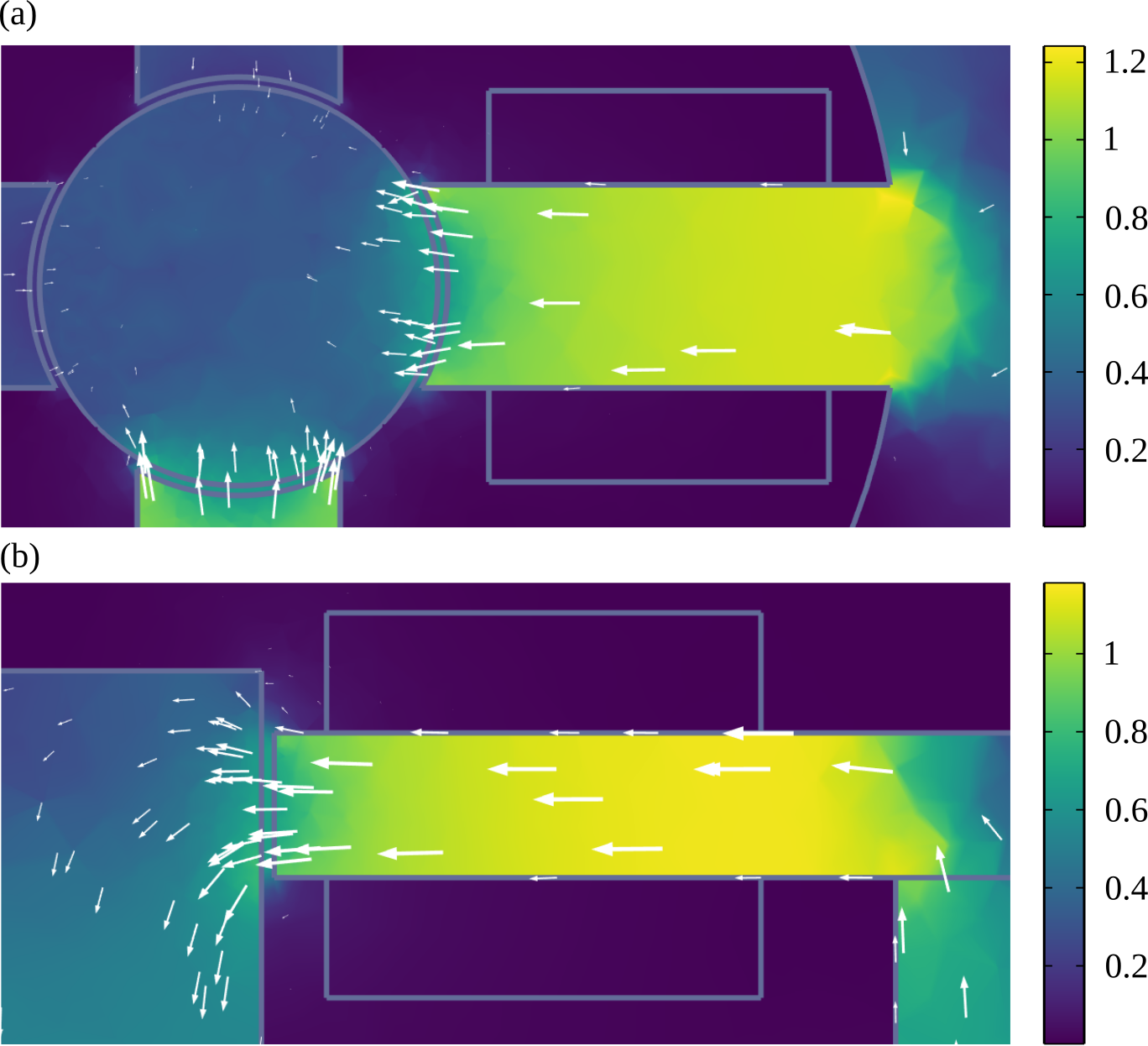}
  \caption{Cross-sections (CS) of AMB (geometry shown in Figure~\ref{fig:RadialAMBCS}) with magnetic flux density in T. The white arrows show the direction of the magnetic flux density field. Maximum airgap flux density stayed around 0.664~T. (a) Transverse CS in mid-stator plane normal to rotor axis. (b) Longitudinal CS in plane parallel to rotor axis.}
  \label{fig:flux}
\end{figure}

As discussed in section~\ref{sec: Topology}, correction factors for AMB analytical model were determined using 3D FE model which takes into account the effects neglected in construction of the analytical model. To verify and attain increased confidence in selected correction factors, the selected design point from the Pareto front was analyzed using 3D FE model to compare with the analytical model. Following the FE model construction explained in section~\ref{sec: Topology}, the magnetic flux densities are plotted for two AMB cross-sections (CS) in Figure~\ref{fig:flux}. The airgap magnetic flux densities stayed at 0.664~T (in airgaps where bias and control flux are in the same direction) and at 0.205~T (in airgaps where bias and control flux are in opposite directions). The radial force was determined at 13.72~N. When compared with results from the analytical model, the maximum relative error in airgap flux densities was 3.4\% while the relative error in radial forces was 1.17\%. Acknowledging that it is challenging to determine correction factors for each combination of AMB geometrical parameters, these errors were considered acceptable to employ corrected analytical models in the optimization model.  \\

\subsubsection{Realization}   \label{sec: RadialAMBRealization}
\begin{figure}[h!]
\centering
  \includegraphics[width=8cm]{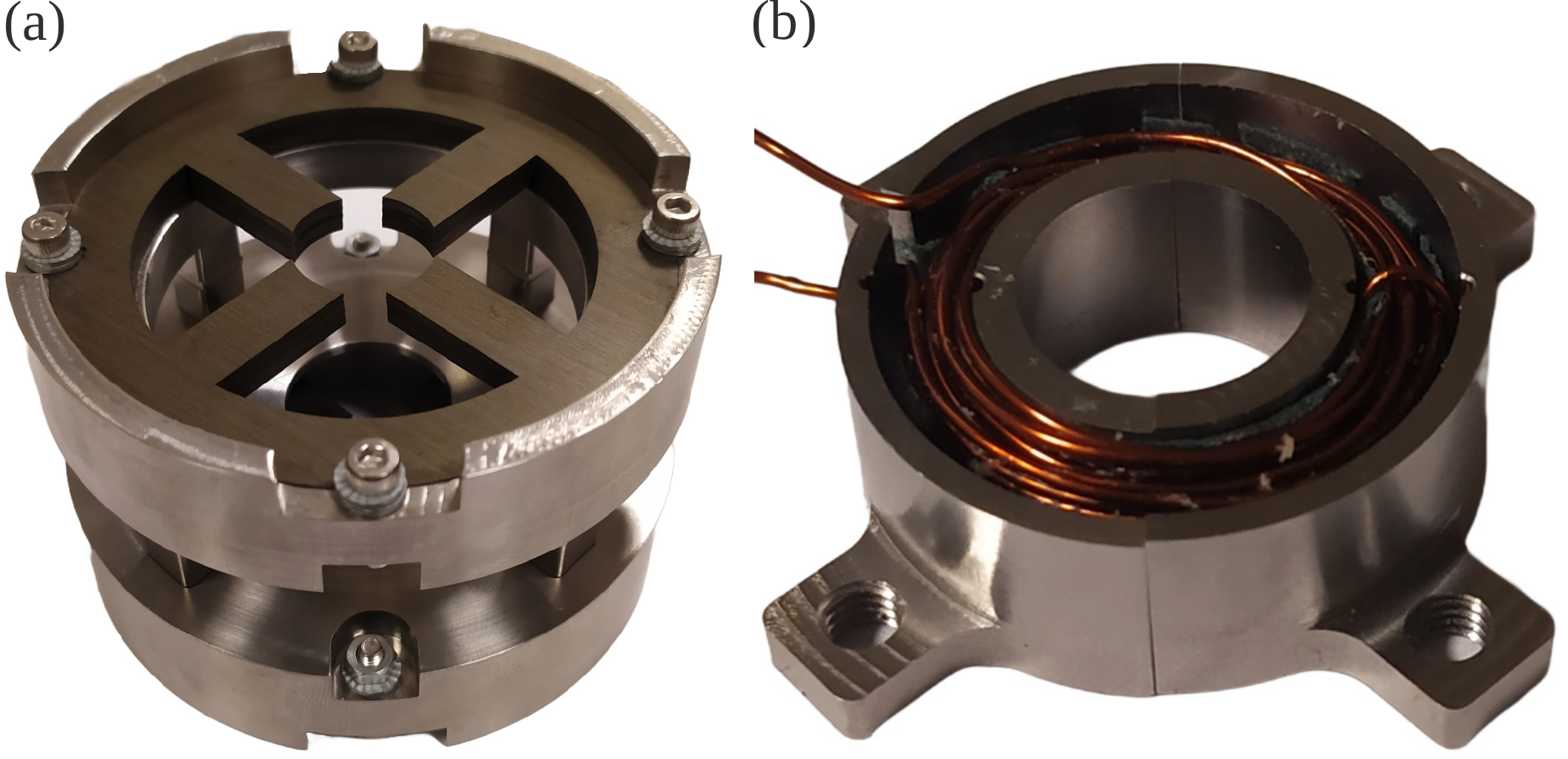}
  \caption{Manufactured AMB sub-assemblies (not to scale). (a) Radial AMB. (b) Axial AMB (lower half; see Figure~\ref{fig:axialAMB}).}
  \label{fig: ManAMB}
\end{figure}

Based on the parameters listed in Table~\ref{tab:DesVar}, radial AMB was realized. For the stator, M270-35A electrical steel was sourced from Somal Sac, Turkiye. Each lamination is 0.35~mm thick. Wire electrical discharge machining was used to cut the lamination stack as it is a burr-free cutting method and minimally impacts the magnetic performance of electrical steel. For AMB coils, we acquired double-coated magnet wire from Emtel, Turkiye. Based on the housing design principles explained in Step 6 of the design framework, radial AMB stators require housing materials with low electrical conductivity to minimize eddy current losses and low magnetic permeability to inhibit undesired flux paths. Austenitic stainless steels provide both properties and hence AISI 316L was acquired. The PM-stator connectors were machined from AISI 430F stainless steel and Neodymium magnets (NdFeB N35) were purchased from Manyet, Turkiye. Realized radial AMB is shown in Figure~\ref{fig: ManAMB}(a).

\subsection{Axial AMB} \label{sec: AxialAMB}

With the four lateral DOFs controlled by the two radial AMBs, the fifth (axial) DOF was controlled by an axial AMB. Following guidelines from Step 4 of the design framework, the design and analysis are presented in this section, along with details on manufacturing.\\

\subsubsection{Topology and modeling} \label{sec: AxialAMB_TopMod}
\begin{figure}[h!]
\centering
  \includegraphics[width=5.5cm]{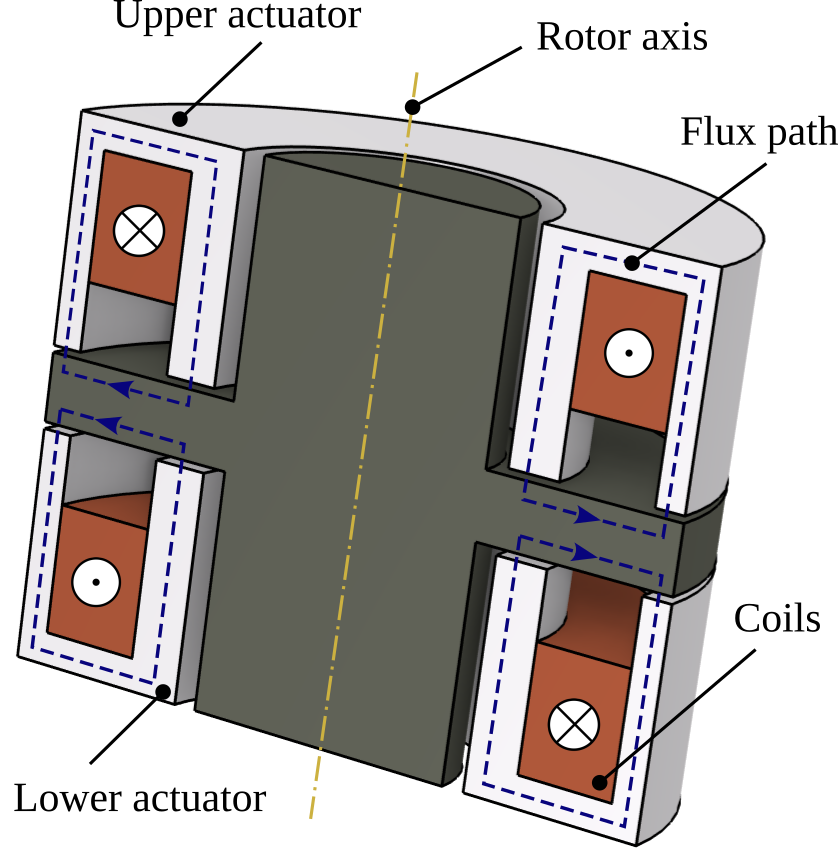}
  \caption{Axial AMB longitudinal cross-section. Two reluctance actuators are placed at either side of the rotor disc for axial levitation control. The dot and cross convention shows the direction of coil currents. Dashed lines show the flux path.}
  \label{fig:axialAMB}
\end{figure}

To levitate the rotor axially and regulate its dynamics, the axial AMB consisting of two circular U-shaped reluctance actuators was used as shown in Figure~\ref{fig:axialAMB} and assembled in Figure~\ref{fig:spindlelabelled}. Placed at each side of the rotor disc, the actuators are driven by tangentially wound coils to produce attractive magnetic forces. The control system commands the coil currents to produce the required net force from the difference of attractive forces from both actuators. The force derivation for axial AMB follows the same assumptions from the radial AMB analytical model (section \ref{sec: Topology}) that flux leakage and magnetic saturation effects were ignored while an infinite permeability was assumed for the core. The net axial force \(F_a\) can be written as \cite{schweitzerMagneticBearingsTheory2009}
\begin{equation}
    F_a = \frac{\mu_0 n_a^2 A_a}{4 k_{ax}^2} \left(\frac{i_1^2}{l_1^2}-\frac{i_2^2}{l_2^2}\right)
    \label{eqn:AxialForce}
\end{equation}
where \(\mu_0\) is the magnetic permeability of air, \(n_a\) is the coil turns per actuator, \(A_a\) is the pole area (each actuator has two poles), \(k_{ax}\) is the correction factor to account for the assumptions taken in the analytical model, \(i_1\) and \(i_2\) are coil currents in upper and lower actuators, respectively, and \(l_1\) and \(l_2\) are upper and lower airgaps, respectively. The correction factor is determined by constructing a 3D FE model for axial AMB similarly as done for radial AMB in section \ref{sec: Topology}. The non-linear (quadratic) dependence of force on coil current creates a challenge as the force-current slope is essentially zero at zero excitation and hence small control currents produce negligible force. Practically, this is addressed by introducing bias flux, provided either by permanent magnets or offset coil current, so that AMB operates at a point with improved force-current gain. This also linearizes the force-current relation at the operating point for linear control design. With this improved force-current gain, a larger force slew-rate is attained with the same current slew-rate, resulting in better AMB dynamic performance. Implementing this approach, the coil currents can be divided into bias current \(i_b\) and control current \(i_c\) as: \(i_1 = i_b + i_c\) and \(i_2 = i_b - i_c\) to produce a net force. Similarly, the airgaps can be written in terms of nominal airgap \(l_o\) (at equilibrium position) and rotor axial displacement \(z\): \(l_1 = l_o - z\) and \(l_2 = l_o + z\). Substituting these current and airgap relations in Eq.~\ref{eqn:AxialForce} and linearizing the force at the operating point (\(i_c=0, z=0\)) using the Taylor series yields
\begin{equation}
    F_a = K_a z + K_{ia} i_c
    \label{Eqn:LinAxialForce}
\end{equation}
where \(K_a\) is the axial force-displacement dependency (also called axial negative stiffness) and \(K_{ia}\) is the force-current gain which are given as
\begin{align}
    K_a &= \frac{\mu_0 n_a^2 A_a i_b^2}{l_o^3 k_{ax}^2}\\
    K_{ia} &= \frac{\mu_0 n_a^2 A_a i_b}{l_o^2 k_{ax}^2}    
\end{align}

Electrical steel laminations provide low eddy current and hysteresis losses in magnetic actuators. However, the mechanical challenges in laminating axial actuators lead designers to use soft magnetic materials which have high magnetic permeability and low coercivity. Therefore, annealed ferritic stainless steel AISI 430F was selected for axial AMB.\\

\subsubsection{Dimensioning and Analysis} \label{sec: AxialAMB_Analysis}
Constrained by spindle geometry, we dimensioned the axial AMB to ensure that load capacities exceed the demands with sufficient margins. The constraints of available coil space dictated by spindle geometry and coil current density for uncooled magnetic actuators were used to fix the maximum coil current. The coil turns per actuator were set at 66 and maximum current was limited to 2.1 A with a bias of 1.6 A. After setting the bias current, a static current offset was added to the upper actuator coils and subtracted from the lower actuator coils to balance the weight of the rotor (2.17~N). The other axial load demand is from the machining forces. From the milling force model in section~\ref{sec:millforces} where a two-flute tool was used, the static (mean) axial force demand stood at 0.77~N and the dynamic demand was 0.25~N at 2$\times$ the rotor frequency. The axial dynamic demand was determined using the same method utilized for radial dynamic demand in section~\ref{sec: Radial-AMB-Analysis}. A safety factor of 2 was applied to the static load and 3 to the dynamic load, consistent with the approach adopted in radial AMB analysis. With the design parameters listed in Table~\ref{tab:AxialAMBParam}, the axial AMB achieved a static load capacity of 5.96~N and a dynamic capacity of 1.47~N at 2$\times$ the rotor frequency (2$\times$1833.33~Hz), exceeding the above-mentioned load demands with sufficient margins. However, similar to radial AMB, the axial AMB faces the same challenge of limited closed-loop bandwidth to actively compensate high-frequency dynamic forces.

\begin{figure}[h!]
\centering
  \includegraphics[width=6cm]{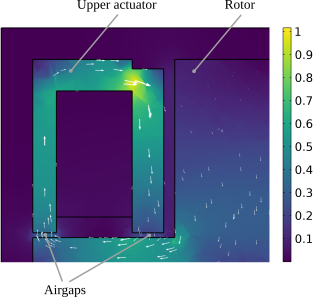}
  \caption{Quarter view of longitudinal cross-section of axial AMB showing the magnetic flux density in T (refer to geometry in Figure~\ref{fig:axialAMB}) when upper actuator coils were excited with a current of 2.1 A. White arrows represent the direction of the flux density field.}
  \label{fig:axialflux}
\end{figure}

In addition to aiding the determination of correction factors, the 3D FE model also identifies locally saturated regions in the flux path. With a maximum current of 2.1 A in upper actuator coils, Figure~\ref{fig:axialflux} shows that only a limited region underwent relatively elevated flux density which still stayed in the linear range (up to 1~T) of the material (AISI 430F) B-H curve and well below the saturation limit.\\

\begin{table}[h]
    \caption{Parameters for axial AMB.}
    \label{tab:AxialAMBParam}
    \centering
    \begin{tabular}{l|l|l}
        \hline
        Bias current                   & $i_b$       & 1.6 A          \\ 
        Axial airgap                   & $l_o$       & 0.3 mm         \\ 
        Stator pole area               & $A_a$       & 1.16$\times$10\(^{-4}\) m\(^2\) \\ 
        Coil turns per actuator        & $n_a$       & 66             \\ 
        Force-displacement dependency  & $K_a$       & 6.4$\times$10\(^{4}\) N/m    \\ 
        Force-current dependency       & $K_{ia}$    & 11.9 N/A       \\ 
        \hline
    \end{tabular}
\end{table}

\subsubsection{Manufacturing}
The axial AMB with parameters mentioned in Table~\ref{tab:AxialAMBParam} was manufactured using standard milling operations. The stator was machined from annealed AISI 430F (Ugitech, France). The spacer plates (see Figure~\ref{fig:spindlelabelled}) supporting the stator were machined with low-conductivity and non-magnetic stainless steel AISI 316L to reduce eddy current losses and prevent unintended flux paths. Double-coated magnet wire (AWG 22 with 200~\textdegree C thermal class) for coils was sourced from Emtel, Turkiye. Holes were drilled in the stators to accommodate non-magnetic binding wires that would hold the coils. Lower half actuator of the manufactured axial AMB is shown in Figure~\ref{fig: ManAMB}(b).

\subsection{Instrumentation and Control System Hardware}
In accordance with Step 4 of the design framework, the required sensing specifications are defined first to select appropriate sensor types. The radial AMB airgap is 0.4~mm and the touchdown clearance is limited to approximately half of this airgap. Hence, the radial displacement sensor standoff and linear operating range must accommodate $\pm0.2$~mm radial motion. Using the same logic, this value for the axial displacement sensor is $\pm0.15$~mm of axial motion. 

For a two-fluted tool, the tooth-passing frequency is 2$\times$ of the spindle rotational frequency. At our operating speed 110,000~rpm (1833.33~Hz), this corresponds to 3.67~kHz. The sensor frequency response should exceed 3.67~kHz with a good margin to avoid significant attenuation and phase lag near the tooth-passing frequency. Adopting a spindle position error budget of 1.5~$\mathrm{\mu m}$ as done by Kimman \textit{et al.}~\cite{kimmanMiniatureMillingSpindle2010}, displacement sensors must have a resolution of less than 1.5~$\mathrm{\mu m}$. Based on 3D FE model in section~\ref{sec: Radial-AMB-Analysis}, the magnetic flux density is only on the order of a few mT at sensor locations between the two radial AMB stators.

Based on these requirements, Sensonics PRI04 eddy current probes were selected as radial and axial displacement sensors. This probe has a measurement range of 2.5~mm, a specified frequency response of DC-10~kHz, and a resolution of less than 1~$\mathrm{\mu m}$. Moreover, the magnetic field effect for this sensor is less than 1\% at 110~mT. Based on electromagnetic FE analysis, the flux density experienced by the probe in our design is significantly lower than 110~mT. For speed measurement, we acquired a Monarch ROLS24-W optical laser sensor which has a specified speed range of 1-250,000~rpm. Control system hardware included dSPACE 1104 and switching power amplifiers. Sensors and control system components can be seen in Figure~\ref{fig:Wholesetup}.

\subsection{Backup Bearing}
\begin{figure}[h!]
\centering
  \includegraphics[width=5cm]{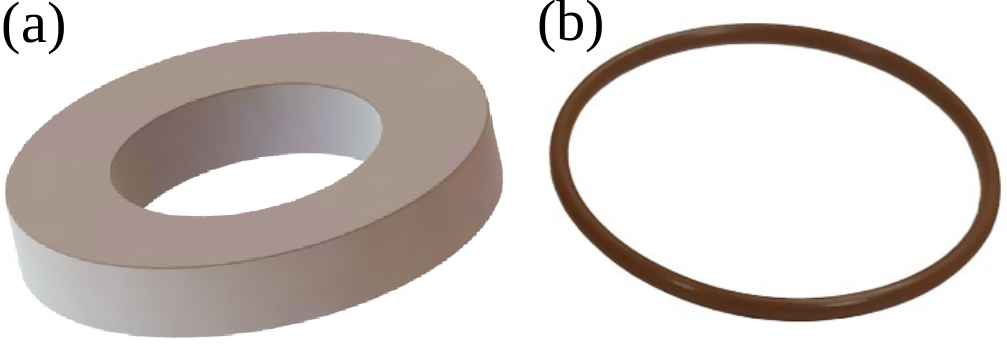}
  \caption{Backup bearing components. (a) Ceramic plain bearing. (b) Viton O-ring.}
  \label{fig:BB}
\end{figure}

Following Step 5 of the design framework, an auxiliary (backup or touchdown) bearing mechanism was incorporated to protect the rotor in the event of AMB failure or excessive displacement. A touchdown clearance of half the AMB airgap was considered appropriate based on common auxiliary bearing design practice~\cite{yuDynamicAnalysisActive2015}. Although ultra-high-speed rolling element bearings are available which can withstand our spindle target speed of 110,000~rpm, a high-speed shock loading requires additional considerations about the stability of internal elements and failure modes. In contrast, plain bearings provide a mechanically simpler and relatively inexpensive solution but a high-friction contact as wear progresses~\cite{karkkainenDynamicSimulationFlexible2007}. Measures such as lubricating or coating the contact interfaces must be taken to reduce friction during touchdown. We opted for ceramic plain bearings as backup bearings. To dampen the touchdown impact on assembly and reduce the possibility of critical behavior, elastomer O-rings were used as a compliant support. Viton O-rings with a shore hardness of 80 were acquired from Eksen Makina, Turkiye. The ceramic plain bearing and Viton O-ring are shown in Figure~\ref{fig:BB} and the placement of the mechanism in the AMB housing is shown in Figure~\ref{fig:spindlelabelled}.

\subsection{Spindle Housing}
\begin{figure}[h!]
\centering
  \includegraphics[width=4cm]{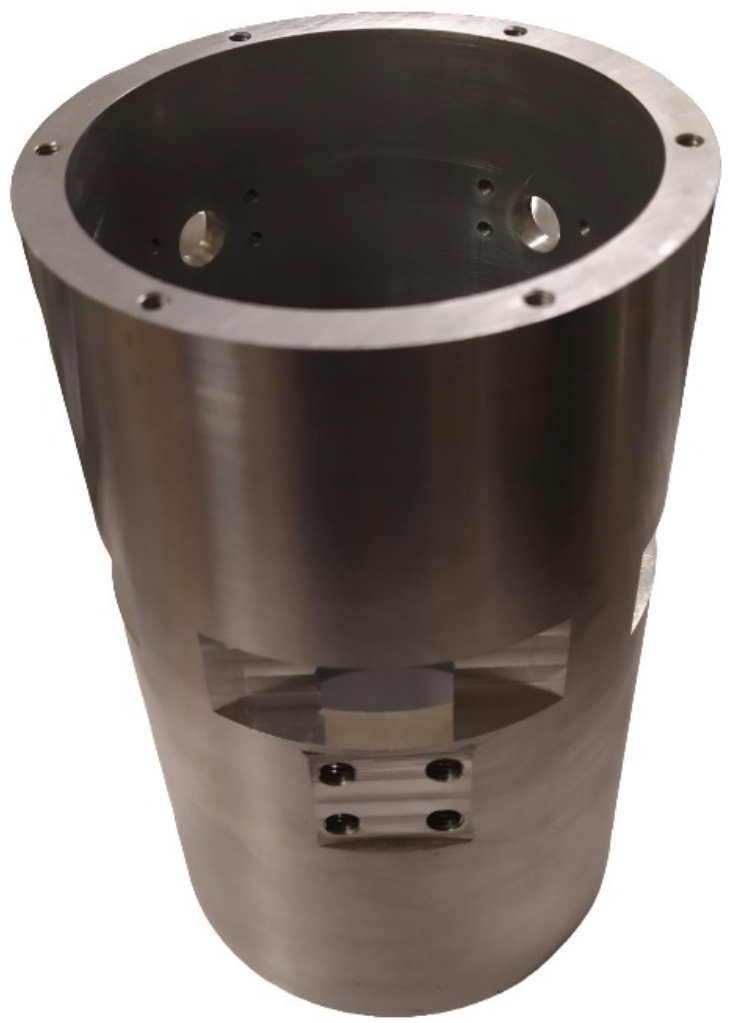}
  \caption{Spindle housing (without end covers).}
  \label{fig:SpindleHousing}
\end{figure}

Referring to Step 6 of the design framework, aluminum was selected as the material for the outer housing for its low mass, high thermal conductivity, and non-magnetic properties. The housing was designed as a stiff protective enclosure to protect AMB components from machining debris. For metals, high thermal conductivity generally comes with high electrical conductivity (Wiedemann-Franz law). Therefore, an aluminum housing may incur significant eddy current losses if exposed to time-varying magnetic fields. However, our design limits the exposure of spindle housing to such magnetic fields and separates this housing from active magnetic circuit by AMB housing. Considering the proximity of spindle housing end covers to AMB coils, covers were machined from low-conductivity and non-magnetic AISI 316L in a similar manner as done for AMB housings. The spindle housing is depicted in Figure~\ref{fig:SpindleHousing}.

\subsection{Thermal Considerations}
From the heat sources mentioned in Step 7 of the design framework, the dominant heat sources in our design are copper losses in AMB coils and core losses in magnetic circuit. Instead of implementing forced cooling, we controlled the temperature rise by limiting the coil current densities to recommended safe ranges for naturally cooled magnetic bearings and motors~\cite{khooSpecificLoadCapacity2007}. Further, we acquired double-coated 200~\textdegree C thermal class magnet wires (compliant with IEC 60317-13) for coils that provide high thermal endurance. Additionally, open slots were machined in the outlet ducts of the nozzle plate as seen in Figure~\ref{fig:AirTurbine}(b). These slots enable a fraction of outlet air to flow axially towards the AMB coils and stators resulting in a limited cooling effect.

\subsection{Assembly} \label{sec: CaseStdyAsm}
The manufacturing and realization of spindle components have been addressed in separate sections of this case study. This section deals with the overall spindle assembly. As per standard practice, a full CAD assembly was prepared to identify critical concentric and axial stack-up dimensions for tighter tolerances and non-critical dimensions where standard machine shop tolerances suffice. The CAD assembly also helped to plan the assembly and disassembly sequences and cable routing.

\begin{figure}[h!]
  \centering
  \includegraphics[width=0.95\columnwidth]{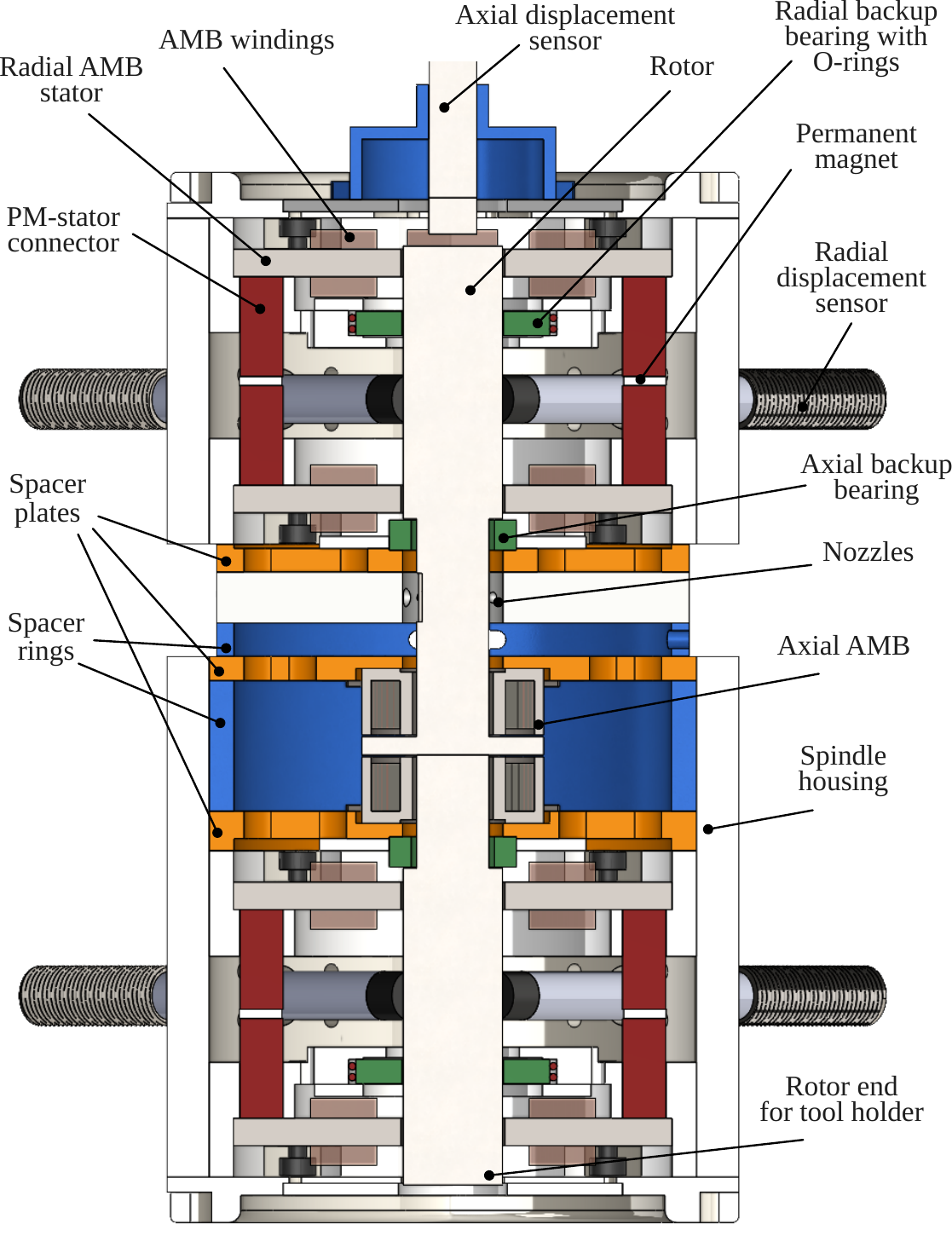}
  \caption{Labeled cross-section of AMB spindle CAD assembly.}
  \label{fig:spindlelabelled}
\end{figure}

\begin{figure*}[h!]
\centering
  \includegraphics[width=14cm]{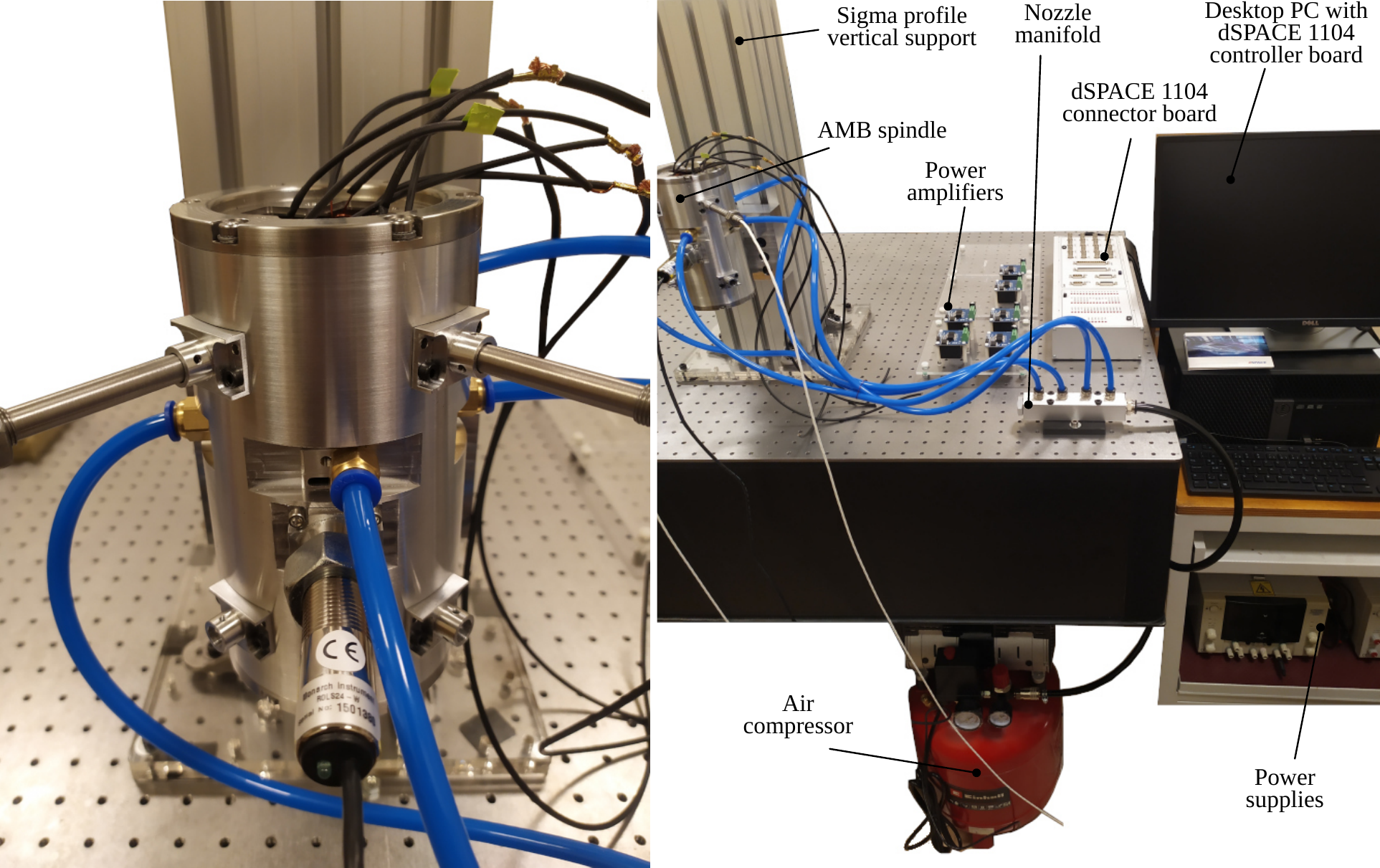}
  \caption{Assembled AMB spindle system.}
  \label{fig:Wholesetup}
\end{figure*}

To secure the radial AMB laminations as a stack and fasten them to the AMB housing, semi-circular slots were machined into the periphery of the lamination stack (via EDM) and the inner bore of AMB housing. Once aligned, these slots form channels for bolted connections. In line with housing material considerations mentioned in section~\ref{sec: RadialAMBRealization}, material for nuts and bolts should also have low electrical conductivity and low magnetic permeability. Therefore, austenitic stainless steel (AISI 304) was selected. To avoid forming an unintended conductive bridge among laminations and consequently increasing the eddy current losses, insulating sleeves separated the bolt from the lamination stack. Locking washers from Nord-Lock were used to prevent self-loosening under vibration loads on the spindle assembly. A CAD cross-section of the spindle is labeled in Figure~\ref{fig:spindlelabelled}. The overall assembled system is shown in Figure~\ref{fig:Wholesetup}.  

\section{Conclusion}  \label{sec: Conc}
Although realizations of AMB spindles and control strategies targeting various applications have been documented in the literature, comprehensive design methodologies for such spindles have not been published. Our study proposed a design framework aimed to unify the design knowledge and demonstrated the utility of the framework by manufacturing a miniaturized micro-milling AMB spindle with a target speed of 110,000~rpm. The multidisciplinary design flow was implemented by considering the mechanical, electromagnetic, and thermal performance of spindle components. The iterative framework incorporated checks for design feasibility, rotordynamic performance, and thermal limits, which may lead to appropriate design iterations. For each component, an explicit focus was applied to its material selection and manufacturing. Future work will focus on control system tuning and micro-milling experiments.



\section*{Acknowledgment}
This work was supported by the Scientific and Technological Research Council of Turkey (TUBITAK) under Grant No. 123M121 (PI: Bekir Bediz).

\bibliographystyle{IEEEtran}
\bibliography{references.bib}




\end{document}